\documentclass[5p,twocolumn,authoryear,twoside,10pt]{elsarticle}

\usepackage{graphicx}
\usepackage{subfigure}
\usepackage{amssymb}
\usepackage{amsmath}
\usepackage{url}
\usepackage{multirow}

\journal{New Astronomy}

  \newcommand{\pad}[2][]{\ensuremath{\frac{\partial #1}{\partial #2}}}

\begin{document}

\begin{frontmatter}

\title{Realistic Simulations of Stellar Surface Convection with ANTARES: \\
I. Boundary Conditions and Model Relaxation}

\author[fmv,mpa]{H.~Grimm-Strele\corref{cor1}}
\ead{hannes.grimm-strele@univie.ac.at} 
\author[fmv]{F.~Kupka} 
\author[fmv]{B.~L\"ow-Baselli}
\author[fmv]{E.~Mundprecht}
\author[tuc]{F.~Zaussinger}
\author[fmv]{P.~Schiansky}

\cortext[cor1]{corresponding author}
\address[fmv]{Institute of Mathematics, University of Vienna, Oskar--Morgenstern--Platz 1,
                       1090~Vienna, Austria}
\address[mpa]{Max--Planck Institute for Astrophysics, Karl--Schwarzschild--Strasse~1, 
                       85741~Garching, Germany}
\address[tuc]{BTU Cottbus, Siemens--Halske--Ring~14, 03046~Cottbus, Germany}

\begin{abstract}
We have implemented open boundary conditions into the ANTARES code to
increase the realism of our simulations of stellar surface convection. Even though 
we greatly benefit from the high accuracy of our fifth order numerical scheme 
(WENO5), the broader stencils needed for the numerical scheme complicate the 
implementation of boundary conditions. 
We show that the effective temperature of a numerical simulation cannot be changed 
by corrections at the lower boundary since the thermal stratification does only change 
on the Kelvin--Helmholtz time scale. Except for very shallow models, this time scale 
cannot be covered by multidimensional simulations due to the enormous computational 
requirements. We demonstrate to what extent numerical simulations of stellar surface 
convection are sensitive to the initial conditions and the boundary conditions. An 
ill-conceived choice of parameters for the boundary conditions can have a severe impact.
Numerical simulations of stellar surface convection will only be (physically) meaningful 
and realistic if the initial model, the extent and position of the simulation box, and 
the parameters from the boundary conditions are chosen adequately.
\end{abstract}


\end{frontmatter}

\section{Introduction}

Nowadays, realistic simulations of stellar surface convection are a mature tool of 
computational astrophysics, and find a wide field of applications. An extensive review 
on this subject is given in \citet{NordlundSteinAsplund2009}. 
\citet{BeeckColletSteffenetal2012} illustrate the usefulness and reliability of results 
obtained from simulations of solar surface convection with the codes CO5BOLD, MURaM and 
Stagger. Despite the differences in numerical schemes and atomic physics between
the simulations they compared, the overall stratification and other basic properties
of the numerical models are similar. 

For any numerical simulation which requires the solution of partial differential 
equations, the choice of the numerical method, the simulation domain, and appropriate 
boundary conditions are of major importance. For the simulation of convection at the 
surface of solar-type stars, the top boundary of the simulation domain is in the 
upper photosphere, and the bottom boundary is situated inside the convective envelope. 
The bottom boundary conditions determine the adiabat of the deep convection zone 
(the part interior to the simulation), and therefore the effective temperature 
$T_{\rm eff}$ of the simulation.

The ANTARES code was originally developed at the University of Vienna for the
simulation of surface convection \citep{MuthsamLoew-BaselliOberscheideretal2007,
MuthsamKupkaLoew-Basellietal2010,LemmererUtzHanslmeieretal2013}. 
Recently, the code was also applied to many other astrophysical problems 
\citep{KupkaBallotMuthsam2009,MuthsamKupkaMundprechtetal2010,KupkaHappenhoferHiguerasKoch2012,
HappenhoferGrimm-StreleKupkaetal2013,ZaussingerSpruit2013,MundprechtMuthsamKupka2013}.
In some of these applications problems arose due to the use of closed boundary conditions
\citep{KupkaBallotMuthsam2009,MundprechtMuthsamKupka2013}. Shock fronts were reflected at 
the top boundary, leading to numerical instabilities.

\citet{RobinsonDemarqueLietal2003}, \citet{KupkaRobinson2007} and \citet{Kupka2009,Kupka2009b} 
discussed, amongst others, the effect of boundary conditions on statistical properties of the 
flow. Unrealistic boundary conditions can lead to unphysical flow patterns in most of 
the simulation box. When the bottom boundary is situated in the convective region of the star, 
which usually is the case, e.g., for solar surface convection simulations, a vast majority
of the energy is transported in the lower part of the simulation box by advection of enthalpy
(the convective flux) and kinetic energy. Therefore, the boundary conditions should
allow transport of energy by convective motions avoiding the need of feeding in the 
energy by an artificial radiative source term. At the bottom boundary, the entropy of 
the inflowing material is unknown and has to be specified somehow. At the top boundary, 
shock fronts should not be reflected, but be transmitted.

Even though a lot of literature exists on stellar surface convection simulations
\citep[e.g.,][]{Nordlund1982,FreytagSteffenLudwigetal2012,VoeglerShelyagSchuessleretal2005,
JacoutotKosovichevWrayMansour2008}, a thorough investigation of the boundary conditions 
effects on the simulations appears to be missing. This paper gives a detailed description 
of the numerical schemes of our surface convection simulations with ANTARES and investigates 
the sensitivity of the simulations to changes in the boundary conditions.

A two-dimensional geometry leads to systematic differences in the mean stratification 
such that, for instance, a different inflow entropy is required to obtain the same total 
flux at the stellar surface \citep{AsplundLudwigNordlundStein2000}. Meaningful tests of 
boundary conditions can ultimately only be done in three dimensions, with sufficient 
resolution and long relaxation time. Despite the nearly linear scaling of our code
\citep{HappenhoferGrimm-StreleKupkaetal2013}, this makes this investigation very 
expensive in terms of computation time as well as in wall-clock time.

The results of the present paper provide a necessary extension to comparisons shown in 
\citet{Kupka2009} and \citet{BeeckColletSteffenetal2012}, where simulations produced with
independent codes were compared. Performing a systematic parameter-space exploration of 
our boundary conditions is simply not affordable. Instead we restrict ourselves to the 
most relevant and fruitful regions of parameter-space. In this paper, we show results 
from several working and non-working combinations and estimate the impact of 
inappropriately designed boundary conditions.

The remainder of the paper is organised as follows. In Section~\ref{section-bc}, we 
describe the details of our implementation of several boundary conditions inspired by 
those used in the CO5BOLD--, MURaM-- and Stagger--codes. The numerical setup of our 
simulations, i.e.\ the starting model and the simulation box size, is described thoroughly 
in Section~\ref{section-domain}. 

We then investigate the effect of using different boundary conditions and compare 
them, taking statistical properties of the flow into account, in Sections~\ref{Res-topBC} 
and~\ref{Res-bottomBC}. 
We show the difficulties arising in the setup of the simulation, its dependence on the initial 
model (Section \ref{initialmodel}) and on the grid resolution (Section \ref{Res-resolution}), 
and emphasize the importance of a careful choice of parameters. In Section~\ref{Res-openclosed}, 
we describe differences between simulations with open and with closed boundary conditions.
Finally, we compare two- and three-dimensional models in Section~\ref{Res-twodim}. 
A discussion of the results can be found in Section \ref{section-discussion} followed by our
conclusions in Section~\ref{section-conclusions}. Finally, in~\ref{app-damp} we 
describe how oscillations can be damped, in~\ref{app-asymmetric} how boundary conditions 
on the gradient are implemented, and in~\ref{app-fluxes} how energy fluxes are calculated.

\section{Boundary Conditions and Initial Models}\label{setup}

In this section, we describe the implementation of several boundary conditions 
in the code ANTARES as well as other modifications to the code we have performed 
in this context.

In this document, the ANTARES convention for spatial coordinates is used. The ANTARES 
code uses a right-handed coordinate system with the $x$--axis pointing into the star. 
$u$ is the vertical component of the velocity vector $(u,v,w)^T$. The grid is equidistant 
in every direction. The methods presented in the following section, however, are not limited 
to an equidistant grid.

The equations we solve are the Navier--Stokes equations

\begin{subequations}
\begin{align}
  \pad[\rho]{t} + \nabla \cdot \left( \rho {\bf u} \right) & = 0, \label{eq-cont} \\
  \pad[\left( \rho {\bf u} \right)]{t} 
    + \nabla \cdot \left( \rho {\bf u} {\bf u} \right) + \nabla p
  & = \rho {\bf g} +  \nabla \cdot \tau, \\
  \pad[E]{t} + \nabla \cdot \left( {\bf u} \left( E + p \right) \right)
  & = \rho \left( {\bf g} \cdot {\bf u} \right) 
    + \nabla \cdot \left( {\bf u} \cdot \tau \right) + Q_{\rm rad},
\end{align}\label{eq-NS}
\end{subequations}

\noindent where ${\bf g}$ is the gravity vector and $Q_{\rm rad}$ is the radiative heating 
rate describing the energy exchange between gas and radiation 
(cf.\ \citealt{MuthsamKupkaLoew-Basellietal2010}). 
The viscous stress tensor $\tau = \left( \tau_{i,j} \right)_{i,j=1,2,3}$ is given by

\begin{equation}
  \tau_{i,j} = \eta \left( \pad[u_i]{x_j} + \pad[u_j]{x_i} 
             - \frac{2}{3} \delta_{i,j} \left( \nabla \cdot {\bf u} \right) \right)
             + \zeta \, \delta_{i,j} \left( \nabla \cdot {\bf u} \right). 
  \label{eq-tau}
\end{equation}

$\delta_{i,j}$ is the Kronecker symbol. $\eta$ and $\zeta$ are the first and second 
coefficients of viscosity. We use the LLNL equation of state from 
\citet{RogersSwensonIglesias1996}, the OPAL opacities from \citet{IglesiasRogers1996} and 
the composition from \citet{GrevesseNoels1993}.
Our standard initial model is {\it model\,S} from \citet{Christensen-Dalsgaardetal1996}
which is calibrated to match helioseismic inversions and uses the same equation of state, 
opacities, and chemical composition.

For the spatial discretisation of the advective part of the Navier--Stokes equations, 
ANTARES uses the weighted essentially non-oscillatory (WENO) algorithm as described, e.g.,
in \citet{JiangShu1996} and \citet{Shu2003}. WENO schemes exist for any order of 
accuracy. In ANTARES, we use the fifth order variant which will be called WENO5 in the 
following \citep{MuthsamKupkaLoew-Basellietal2010}, without any artificial stabilisation 
terms. The advantages of the WENO5 method are its high order of accuracy in smooth flow regions 
and its shock-capturing ability. This allows accurate modelling of surface convection 
where strong shock fronts are ubiquitous in the photosphere, whereas the granules themselves 
are rather smooth (cf.\ \citealt{NordlundSteinAsplund2009}). The appropriate treatment of 
diffusive and viscous terms in the context of WENO5 is discussed in detail in 
\citet{HappenhoferGrimm-StreleKupkaetal2013}.

The high accuracy of a fifth-order scheme, combined with the stability of WENO into the 
WENO5 algorithm used in ANTARES as shown in \citet{MuthsamLoew-BaselliOberscheideretal2007} 
and \citet{MuthsamKupkaLoew-Basellietal2010} justifies the use of broader stencils needed 
for any high-order method. In the design of the boundary conditions, we must 
keep in mind that the boundary extends over several vertical layers.

\subsection{Boundary Conditions for Stellar Convection Simulations}\label{section-bc}

For solar-like main-sequence stars, the characteristic scale of surface convection is 
small compared to the stellar radius and, hence, the ``box--in--a--star'' approximation is 
valid for the simulation of surface convection. When the turbulent kinetic energy
of the simulation is unaffected by a further increase in aspect ratio of the domain,
periodic boundary conditions can be used at the horizontal boundaries
\citep{RobinsonDemarqueLietal2003}.

In the vertical direction, the top boundary is typically placed 
above the photosphere where\-as the bottom boundary is in the upper part of 
the convective envelope \citep{SteinNordlund1998,RobinsonDemarqueLietal2003,
WedemeyerFreytagSteffenLudwigHolweger2004,FreytagSteffenLudwigetal2012,
VoeglerShelyagSchuessleretal2005,
MuthsamLoew-BaselliOberscheideretal2007,MuthsamKupkaLoew-Basellietal2010}. 
Numerical boundary conditions either force the fluid to stop its vertical motion 
(the so called closed boundary conditions) and hence not cross the vertical boundary, 
or they allow in-­ and outflow (open boundary conditions). In each of these two classes, 
there are many variants available for the design of specific boundary conditions.

Simulations from \citet{SteinNordlundGeorgovianietal2009} show that the dominant 
horizontal scale of the convection increases monotonically with increasing depth 
in the uppermost $\sim 20\,{\rm Mm}$ of the solar convection zone, whereas the area 
ratio of upflows to downflows is nearly independent of depth, and the average energy 
fluxes stay nearly constant. There is no physical reason why the vertical convective 
motions should stop at the boundary of the simulation domain situated significantly
above the boundary of their simulations as it is forced by closed boundary conditions. 
To increase the physical realism of the simulation it is mandatory to allow free in- and 
outflow at the boundaries. But the design of stable and benign open boundary conditions 
is non-trivial, as we show in the following. Since we intend to use symmetric stencils 
at the boundary, the boundary conditions must be set on several layers corresponding to 
the width of the stencils.

\subsubsection{Closed Boundaries}\label{closedbc}

Slip (stress-free) boundary conditions are characterised by setting the vertical velocity 
component and the vertical derivative of the horizontal velocities to $0$, i.e.\

\begin{subequations}
\begin{equation}
  u = 0,\ \frac{\partial v}{\partial x} = \frac{\partial w}{\partial x} = 0.
\end{equation}

Furthermore, since the bottom boundary can transport energy only by radiation, an 
artificial source term must be introduced to feed the required amount of energy into 
the simulation domain. This is done by modifying the thermal conductivity $k$ 
in the lowermost layers of the simulation domain such that 

\begin{equation}
  - k_{{\rm modified}} \frac{\partial T}{\partial x} = F_{\star} = \sigma T_{\rm eff}^4,\label{eqsource}
\end{equation}\label{eq-closed}
\end{subequations}

\noindent and imposing $\frac{\partial \rho}{\partial x} = 0$. The detailed scheme
is described, e.g., in \citet{RobinsonDemarqueLietal2003} and 
\citet{MuthsamKupkaLoew-Basellietal2010}.

Closed boundary conditions have the advantage of simplicity and high 
stability, at least for subsonic flows. On the other hand, they reflect shocks 
and disturb the velocity field in undesirable ways, as discussed by 
\citet{RobinsonDemarqueLietal2003}, \citet{KupkaBallotMuthsam2009}, and 
\citet{Kupka2009}. Effects on the thermodynamics are indirect at most, e.g.\ 
through the reflection of shocks. Nevertheless, there are applications where the use 
of closed boundary conditions is not problematic, such as the numerical simulation of 
the superadiabatic layer (SAL in the following) of the Sun 
\citep{KimChan1998,RobinsonDemarqueLietal2003,KupkaRobinson2007,Kupka2009,
TannerBasuDemarque2012,TannerBasuDemarque2013}. 
Moreover, for hotter main-sequence stars, e.g., the surface convection zone is 
shallow such that the whole convection zone can be contained in the simulation 
domain, and the lower boundary is in the underlying radiative zone. On the other 
hand, the top boundary should be open in their case, too, due to the ubiquity of shocks.

\subsubsection{Open Top Boundary Conditions}\label{topbc}

The top boundary of simulations of surface convection typically is located 
in the photosphere, i.e.\ the region where the optical Rosseland depth $\tau$
is between $10^{-6}$ and $10$. We implemented the top boundary conditions presented in 
\citet{Cheung2006} and \citet{Trampedach1997} with some slight modifications which 
will be described in the following.
Generally speaking, the boundary conditions should be designed in such a way that they 
do not influence the behaviour of the flow. For the top boundary layers, we assume a 
hydrostatic and isoenergetic stratification. For the Sun and the LLNL equation of 
state, this implies nearly isothermal stratification. As a difference to previous 
implementations, in the hydrostatic equilibrium we want to include the turbulent pressure 
since it amounts to $\sim 10\,\%$ of the total mean pressure in the photosphere, which 
is of similar magnitude as in the SAL (the maximum of $\sim 15\,\%$ is reached 
around $100\,{\rm km}$ below the stellar surface). The velocities are constantly 
extrapolated according to

\begin{subequations}
\begin{equation}
  \frac{\partial u}{\partial x} = \frac{\partial v}{\partial x} = \frac{\partial w}{\partial x} = 0
  \label{veltop}
\end{equation}

\noindent to make the boundary transmissive for waves. In order to use the symmetric stencils 
of the fifth order WENO method at the boundary, the boundary conditions must extend over
three fiducial layers.

We assign the indices $-2$, $-1$ and $0$ to the fiducial layers and number the domain
layers starting with $1$ at the top of the domain. The mean density in layer $0$, 
$\langle \rho \rangle_0$, is advanced in time by the continuity equation~\eqref{eq-cont}
setting the mean mass flux from the top to $0$, i.e.\ 

\begin{equation}
  \langle \rho \rangle_0^{(n+1)} = \langle \rho \rangle_0^{(n)} 
        - \frac{{\Delta t}_{\rm stg}}{\Delta x} \langle \rho u \rangle_{\frac{1}{2}}^{(n)},
  \label{eq-rhoupd}
\end{equation}

\noindent where ${\Delta t}_{\rm stg}$ is the time step size of the current Runge--Kutta 
stage and $\Delta x$ is the (constant) vertical grid spacing. $n$ is the number of 
the current time step. $\langle \rho u \rangle_{\frac{1}{2}}^{(n)}$, the mean density 
flux at the mid-point between the layers~$0$ and~$1$, can be computed using the standard
symmetric stencils. With the mean density in layer~$0$ obtained using~\eqref{eq-rhoupd}, 
the density profile layer in~$0$ is just copied from layer~$1$ scaled by the ratio 
$\frac{\langle \rho \rangle_0^{(n+1)}}{ \langle \rho \rangle_1^{(n+1)}}$, i.e.\ 

\begin{equation}
  \rho_0^{(n+1)}(y,z) = \frac{\langle \rho \rangle_0^{(n+1)}}{\langle \rho \rangle_1^{(n+1)}} \rho_1^{(n+1)}(y,z).
\end{equation}
   
The specific internal energy $\langle \epsilon \rangle_0$ in all boundary layers is assumed to 
be constant in space. It is initialised by the mean specific internal energy in the outermost 
domain layer $\langle \epsilon \rangle_1$ and is furthermore advanced in time by

\begin{equation}
  \langle \epsilon \rangle_0^{(n+1)} = (1.0 - \delta) \langle \epsilon \rangle_0^{(n)} 
                                            + \delta \langle \epsilon \rangle_1^{(n+1)}.
  \label{eq-epsilon0}
\end{equation}

\citet{Cheung2006} set $\delta$ constantly to $10^{-3}$, but we reset 
$\langle \epsilon \rangle_0$ only at the end of each Runge--Kutta step with 
$\delta$ given by the formula

\begin{equation}
  \delta = \frac{\Delta t \cdot \langle v_{\rm snd} \rangle_1 }{c_{\rm f} \cdot \langle H \rangle_1},
  \label{eqdelta}
\end{equation}

\noindent where $\Delta t$ is the time step of one Runge-Kutta integration, 
$\langle v_{\rm snd} \rangle_1$ is the mean sound speed in layer $1$ and 
$\langle H \rangle_1$ the mean pressure scale height in that layer. With
the parameter $c_{\rm f}$ we can control $\delta \in [0,1]$, i.e.\ the 
time scale on which the heat content of the boundary layers varies. Higher 
values of $c_{\rm f}$ lead to slower changes. By using $\langle H \rangle_1$
instead of the geometrical height of the computational cell, we keep $\delta$
unchanged when the vertical resolution of the simulation is changed. 
Since $c_{\rm f}$ has to be adjusted anyway, it is sufficient to use the 
gas pressure to compute $\langle H \rangle_1$ for the approximate time scale
underlying \eqref{eqdelta}.

The idea behind equation \eqref{eqdelta} is to make the procedure independent 
of different time integration methods and spatial resolution. With a constant 
value of $\delta$, the time scale of changing $\langle \epsilon \rangle_0$ would 
be different, if a two- or a three-stage Runge--Kutta method is used, or if 
the Courant number is changed. Furthermore, we choose $\delta$ independent 
of the grid spacing such that the time scale does not change if the grid 
is modified. Modifying $c_{\rm f}$ allows to control the stability and the 
stratification of the boundary layers.

The density in the layers above index $0$ is assumed to be pointwisely
in hydrostatic equilibrium

\begin{equation}
  \frac{\partial (p+p_{\rm turb})}{\partial x} = - \rho g,\label{modhydstat}
\end{equation}

\noindent where $p = p_{\rm rad}+p_{\rm gas}$ and $p_{\rm turb}
                   = \rho \left( u-\langle u \rangle \right)^2$ 
is the turbulent pressure. By the chain rule and due to equation~\eqref{veltop},

\begin{equation}
  \frac{\partial (p+p_{\rm turb})}{\partial x} 
    = \frac{\partial \left( p+p_{\rm turb} \right)}{\partial \rho} \frac{\partial \rho}{\partial x}
    = \left( \frac{\partial p}{\partial \rho} 
    + \left( u-\langle u \rangle \right)^2 \right) \frac{\partial \rho}{\partial x}.\label{eqPbound}
\end{equation}

For a perfect gas, constant $\epsilon$ implies an isothermal stratification. The slight
deviations of $T$, evaluated from the LLNL equation of state, from its horizontal mean 
therefore show the deviation from the perfect gas equation.
Since in our simulations, these deviations rarely exceed $5\,\%$, it is 
reasonable to assume the simplification of a perfect gas in the boundary layers, considering the 
other simplifications like constant velocity and constant $\epsilon$. In this case, 
$\frac{\partial p}{\partial \rho} = \left( \gamma - 1 \right) \epsilon$ is constant in 
the boundary layer. 

Since $\left( u-\langle u \rangle \right)^2$ in the boundary layer is constant in the vertical direction due 
to~\eqref{veltop}, the value $c_{\rho}:=\frac{\partial p}{\partial \rho}+\left( u-\langle u \rangle \right)^2$ 
is constant vertically, but varies horizontally. Therefore, we can integrate equation 
\eqref{eqPbound} analytically and obtain

\begin{equation}
  \rho(i,y,z) = \rho(0,y,z) \exp\left( \frac{- |i| \Delta x \cdot g}{c_{\rho}} \right),   \label{rhotop}
\end{equation}\label{eq-topbc}
\end{subequations}

\noindent where $i=-1,-2$.
With $\langle \epsilon \rangle_0$ and the velocities calculated by \eqref{veltop}, 
all values in all boundary layers can be calculated.

\subsubsection{Open Bottom Boundary Conditions}\label{bottombc}

We formulate two sets of boundary conditions, the first one based on 
\citet{NordlundStein1990} and \citet{FreytagSteffenLudwigetal2012}, and the 
second one on \citet{VoeglerShelyagSchuessleretal2005}.

\citet{NordlundStein1990} and \citet{FreytagSteffenLudwigetal2012} set density 
and internal energy such that the inflows have a certain entropy value. This 
entropy value is fixed in space and time and found by trial and error or 
by using previous simulations (see Fig.~3 in \citealt{LudwigFreytagSteffen1999}
and Fig.~1 in \citealt{TrampedachAsplundColletNordlundStein2013}). Finally,
the velocities in the outflows are scaled to obtain an average mass flux of $0$
at the lower boundary.

Instead of prescribing an entropy value, \citet{VoeglerShelyagSchuessleretal2005}
chose to set the specific internal energy of an inflow through a feed-back loop. 
Since in a relaxed simulation, the radiative flux at the top of the domain corresponds to 
the amount of energy flowing through the bottom boundary, \citet{VoeglerShelyagSchuessleretal2005} 
decided to directly correlate the two values on the Kelvin--Helmholtz time scale. When the 
radiative flux at the top is too high, the amount of energy flowing in at the bottom is 
lowered which in principle leads to a lower radiative flux. This motivated the formula

\begin{equation}
  \epsilon_{\rm inflow}^{(n+1)} = \epsilon_{\rm inflow}^{(n)} \cdot \left[ 1.0 + \frac{\Delta t}{\tau_{\rm KH}}
                                        \left( 1.0 - \frac{F_{\rm rad}^{\rm top}}{F_{\star}} \right) \right],
                                                       \label{epsFrad}
\end{equation}

\noindent where $F_{\star} = \sigma T_{\rm eff}^4$ is the energy flux of the star 
according to the Stefan--Boltzmann law and $\Delta t$ the time step size of one 
Runge--Kutta integration. The Kelvin--Helmholtz time scale $\tau_{\rm KH}$ can be 
calculated as

\begin{equation}
  \tau_{\rm KH} = \frac{\int_{{\rm box}} \rho \epsilon\,dV}{\int_{\rm top} F_{\rm rad}\,dy\,dz}   \label{eq_tau_KH}
\end{equation}

\noindent (cf.\ formula (19) and (20) in \citealt{VoeglerShelyagSchuessleretal2005}).
They correct the density at the lower boundary to keep the total mass in the 
simulation domain unaltered. The entropy of an inflow will vary with the changing 
density.
 
We implemented both approaches which define either $S_{\rm inflow}$ or 
$\epsilon_{\rm inflow}$ into our code and show results in Section \ref{Res-bottomBC}. 
In those cases where we specify $S_{\rm inflow}$ at the bottom, we use an iterative 
correction procedure similar to \eqref{epsFrad}, which is used for $\epsilon_{\rm inflow}$.
In the first variant, we only change the time scale and define

\begin{equation}
  S_{\rm inflow}^{(n+1)} = S_{\rm inflow}^{(n)} \cdot \left[ 1.0 + \frac{\Delta t}{\tau_{\rm S}} 
                                        \left( 1.0 - \frac{F_{\rm rad}^{\rm top}}{F_{\star}} \right) \right],
                                                       \label{SFrad}
\end{equation}

\noindent where $\tau_{\rm S}$ is the time scale of the correction. In general, 
$\tau_{\rm S} \neq \tau_{\rm KH}$ since we expect much smaller changes in $S$ 
than in $\epsilon$. In the limits $\tau_{\rm S} \to \infty$ and $F_{\rm rad}^{\rm top} 
\to F_{\star}$, $S_{\rm inflow}={\rm constant}$ as
in \citet{NordlundStein1990} and \citet{FreytagSteffenLudwigetal2012}.

In the second variant, it is guaranteed that the total flux at the bottom boundary is 
close to $F_{\star}$. Here,

\begin{equation}
  S_{\rm inflow}^{(n+1)} = S_{\rm inflow}^{(n)} \cdot \left[ 1.0 + \frac{\Delta t}{\tau_{\rm S}} 
                                        \left( 1.0 - \frac{F_{\rm tot}^{\rm bot}}{F_{\star}} \right) \right],
                                                       \label{SFtot}
\end{equation}

\noindent where $F_{\rm tot}^{\rm bot}$ is the sum of the radiative, 
the kinetic, and the convective fluxes at the bottom of the domain.
In this case, $\tau_{\rm S}$ must be set according to the time scale 
on which $F_{\rm tot}^{\rm bot}$ changes to avoid decoupling of the 
correction mechanism from the correction quantity.

None of these correction mechanisms should be applied during transients 
excited by any changes such as scaling of variables of the simulation. Both the 
radiative flux at the top and the total flux at the bottom require some time to 
reach the value which reflects the current thermal stratification.
Therefore, we do not change $S_{\rm inflow}$ during the first five sound 
crossing times of a new model started from a one-dimensional stratification.

In the following, we describe our implementation of the boundary conditions
following \citet{FreytagSteffenLudwigetal2012} in detail. We assume the velocities 
to be constant with depth

\begin{equation}
  \frac{\partial u}{\partial x} = \frac{\partial v}{\partial x} = \frac{\partial w}{\partial x} = 0.
  \label{velbot}
\end{equation} 

Then, we calculate the density $\rho$ and the specific internal energy $\epsilon$
in the boundary layers. Superscript numbers in paranthesis denote different correction 
levels. In the first correction step of \citet{FreytagSteffenLudwigetal2012}, $\rho$
and $\epsilon$ are reset by

\begin{subequations}
\begin{align}
      \rho^{(1)} & = \rho     + \delta_S \frac{\Delta t}{t_{\rm char}} 
                         \frac{-\rho^2 T (\Gamma_3 - 1)}{p \Gamma_1}      \left( S_{\rm inflow} - S \right), \\
  \epsilon^{(1)} & = \epsilon + \delta_S \frac{\Delta t}{t_{\rm char}} 
                         T \left( 1 - \frac{\Gamma_3-1}{\Gamma_1} \right) \left( S_{\rm inflow} - S \right),
\end{align}

\noindent (cf.\ (34) and (35) from \citealt{FreytagSteffenLudwigetal2012})
with the parameter $\delta_S$ controlling the time scale of the 
correction step. $\Delta t$ is the time step size of the (Runge--Kutta) integration 
and the characteristic time-scale for the adjustments is given by 

\begin{equation}
  t_{\rm char} = \frac{\Delta x}{\langle v_{\rm snd}+\left|u\right|\rangle}
\end{equation}
\end{subequations}

\noindent (cf.\ (33) from \citealt{FreytagSteffenLudwigetal2012}). The adiabatic
sound speed $v_{\rm snd}$ as well as the adiabatic coefficients $\Gamma_1$ and
$\Gamma_3$ are evaluated from the equation of state as a function of the 
uncorrected values of $\rho$ and $\epsilon$.  

In our implementation, however, we obtain $\epsilon^{(1)}$ and $\rho^{(1)}$ 
from performing an isobaric entropy change, i.e.\ we invert the equation of 
state such that

\begin{equation}
  S(\rho^{(1)},\epsilon^{(1)}) = S_{\rm inflow}^{(n+1)},  \label{Seq1}
\end{equation}

\noindent without changing $p$.
$S_{\rm inflow}^{(n+1)}$ is determined either with~\eqref{SFrad} or~\eqref{SFtot}. 
Entropy, internal energy, and density of an outflow are not changed. In this way, 
we delete the parameter $\delta_S$ from the formulation of the boundary conditions.

We then proceed with two additional correction steps proposed 
by \citet{FreytagSteffenLudwigetal2012} which avoid the generation of unwanted
large pressure fluctuations by the inflow at the open bottom boundary, and fix 
the mean mass flux over the boundary to $0$. Thus, as in the second correction 
step from \citet{FreytagSteffenLudwigetal2012}, density and specific internal 
energy throughout the entire boundary layers are reset by

\begin{subequations}
\begin{align}
      \rho^{(2)} & =     \rho^{(1)} + \delta_p \frac{\Delta t}{t_{\rm char}} 
                         \frac{1}{v_{\rm snd}^2} \left( \langle p \rangle - \, p \right),     \label{eq_rho2}    \\
  \epsilon^{(2)} & = \epsilon^{(1)} + \delta_p \frac{\Delta t}{t_{\rm char}} 
                         \frac{1}{\Gamma_1 \rho} \left( \langle p \rangle - \, p \right)
\end{align}
 
\noindent (cf.\ (36) and (37) from \citealt{FreytagSteffenLudwigetal2012}).
The parameter $\delta_p$ governs the time scale on which pressure 
fluctuations are reduced. 

Finally, $\rho$ and the vertical velocity $u$ are modified by

\begin{align}
  \rho^{(3)} & = \rho^{(2)} + \langle \rho \rangle^{(0)} - \langle \rho^{(2)} \rangle,      \label{eq_rho3}   \\
     u^{(1)} & = u - \frac{\langle\rho^{(3)}u\rangle}{\langle\rho\rangle^{(0)}}
\end{align}\label{eq-co5bold}
\end{subequations}

\noindent (cf.\ (38) and (39) from \citealt{FreytagSteffenLudwigetal2012})
to keep the total mass at the bottom boundary unaltered. Again, this
correction is applied throughout the entire boundary layers.

As for the top boundary at the bottom three boundary layers are needed for the WENO scheme
used in ANTARES. After the procedure described above is applied to the innermost boundary 
layer at the bottom, the two underlying layers are filled by exponentially extrapolating 
the density and by linearly extrapolating the specific internal energy (assuming
adiabatic stratification instead would require to proceed as in Sect.~\ref{Sect_DeepModels}
to recover the power law growth with depth, possibly with a correction for turbulent pressure 
as in \eqref{modhydstat} and \eqref{eqPbound}, but we expect the difference to our simplified 
procedure along the two outer grid points to be acceptably small).

When implementing the boundary conditions which are similar to \citet{FreytagSteffenLudwigetal2012}, 
care must be taken in the formulation of mass conservation. Instead of the variable
$\rho u$, the numerical density flux situated at the half-integer node must be set to
$0$. The system is not overdetermined when we enforce this condition, since it is only 
another numerical consequence of the (analytical) mass conservation requirement.

Furthermore, as another available alternative to the open bottom boundary conditions
specified by~\eqref{SFrad} or~\eqref{SFtot}, \eqref{velbot}, \eqref{Seq1}, and 
\eqref{eq-co5bold}, we implemented the full version of boundary conditions 
from~\citet{VoeglerShelyagSchuessleretal2005} into our code. For the present paper 
these are used only for the case of Model~2 in Section~\ref{Res-bottomBC}. In these 
boundary conditions, in addition to specifying the inflow internal energy 
by~\eqref{epsFrad}, the pressure is assumed to be uniform across the lower boundary. 
Its value $p_{\rm tot}$ is determined to keep the mass in the simulation box unaltered 
(see equations~(21) to (23) in the cited reference).In inflow regions, 
\citet{VoeglerShelyagSchuessleretal2005} force the inflow to be vertical (cf.\ (18) 
therein). In contrast, the velocities are set by~\eqref{velbot} in our 
implementation. In downflow regions, the velocities and the entropy density $S$ are 
set according to

\begin{equation}
    \frac{\partial u}{\partial x} = \frac{\partial v}{\partial x} 
  = \frac{\partial w}{\partial x} = 0,\,
    \frac{\partial S}{\partial x} = 0,   \label{eq-muram}
\end{equation}  

\noindent (see equations~(16) and (17) in the same reference).

Open bottom boundaries can lead to a strong increase in horizontal momenta since, due 
to \eqref{velbot}, they do not conserve mass and momentum simultaneously. To mitigate 
this effect we usually damp the horizontal momenta  with the procedure described 
in~\ref{app-damp} in the layers where the bottom boundary condition is applied.

Table~\ref{tab-botBC} provides a summary of models and boundary conditions which we
test in Section~\ref{Res-bottomBC}.

Our standard initial model {\it model\,S} \citep{Christensen-Dalsgaardetal1996} 
has been calibrated to give the right effective temperature for the Sun. Moreover, 
using the composition from \citet{GrevesseNoels1993}, it also gives the same depth 
of the convection zone as obtained from helioseismology. This implies that if we use 
the entropy value from {\it model\,S} for $S_{\rm inflow}$ and the same chemical 
composition as well as opacity and EOS data, we can expect that the numerical simulation 
gives the correct effective temperature. Nevertheless, correction mechanisms 
like~\eqref{epsFrad}, \eqref{SFrad} or~\eqref{SFtot} can change $S_{\rm inflow}$ for 
several reasons: to test if adjustments at the lower boundary can correct small errors
in $T_{\rm eff}$ at the upper boundary \citep{VoeglerShelyagSchuessleretal2005} 
or to control the total energy flux at the lower boundary (cf.\ equation~\eqref{SFtot}).

\subsection{Simulation Domain Size and Initial Conditions}\label{section-domain}

Numerical simulations of stellar surface convection are commonly initialised with 
one-dimensional models taken from stellar evolution calculations. These models 
often do not extend high enough into the atmosphere or exhibit other 
deficiencies as, e.g., unrealistic temperature profiles due to the use of 
the diffusion approximation, which do not allow to apply them directly. In general 
we have to extend the starting models both at the top and the bottom. In the 
following, we describe the procedures used for this purpose.

\subsubsection{Extending the Simulation Box in the Photosphere}

To decrease the influence of the top boundary on the simulation, it is desirable to 
move the top boundary away from the stellar surface towards the upper photosphere 
or even further above. In this paper, the term ``stellar surface'' designates 
the position where the mean temperature equals the effective temperature. Since the 
one-dimensional models used as an initial condition often do not contain data for 
sufficiently low optical depth and moreover since turbulent pressure and multidimensional
radiative transfer cause an elevation of the stellar surface in comparison with 
one-dimensional models \citep{NordlundStein1999,RosenthalChristensen-DalsgaardNordlundSteinTrampedach1999},
we developed a procedure to extend the simulation domain when setting up the initial
conditions from such models.

For a given extension height, the pressure $p$ is exponentially extrapolated 
from the values in the initial model. Then, the density $\rho$ is set according to 
the equation of hydrostatic equilibrium

\begin{equation}
  \frac{\partial p}{\partial x} = - \rho g,
\end{equation}

\noindent where the derivative of $p$ is calculated numerically with fourth order 
centred differences (a negative sign for $g$ as $x$ increases with depth).
By calling the equation of state, $T=T(p,\rho)$ is calculated. With the new 
values of $T$ and $\rho$ the corresponding pressure values are calculated again and the 
whole procedure is repeated several times to ensure consistency with the equation of 
state. The velocities are then set according to~\eqref{veltop}.

In the end, the initial model for the new layers is only of minor importance, since 
due to the short time scales in the upper photosphere, the state variables change 
rapidly away from the initial values. It is indeed more important that this procedure 
is robust than it is an accurate physical model for those layers.

\subsubsection{Deep Models}   \label{Sect_DeepModels}

When we tried to calculate deeper solar granulation models which extend to 
a depth of about $4\,{\rm Mm}$ below the surface, we experienced stability 
problems when the values from the one-dimensional starting model were used for 
initialisation of the state variables. This probably stems from the fact that 
the 3D simulation obtained by interpolating the one-dimensional profile 
to the numerical grid is not numerically adiabatic and hydrostatic.
The deviations from equilibrium and the ensuing instabilities get larger the 
deeper the model is. \citet{RobinsonDemarqueLietal2003} experienced similar 
problems.

Therefore, in the solar case we re-integrate such initial models starting from 
around $2.5\,{\rm Mm}$ below the surface going inwards into the star using 
the classical, explicit 4$^{\rm th}$ order Runge--Kutta algorithm to solve the 
equations of adiabaticity and hydrostatic equilibrium, i.e.\

\begin{align}
  \frac{\partial p}{\partial x} = - \rho g, \\
  \frac{\partial T}{\partial x} = - \rho g \frac{T}{p}\nabla_{\rm ad},
\end{align}

\noindent with $\nabla_{\rm ad}$ and $\rho$ for given $p$ and $T$ taken from the 
equation of state.

Since we have assumed the turbulent pressure to be negligibly small here,
the open bottom boundary has to be located sufficiently deep below the maximum 
of the $p_{\rm turb}/p$--ratio which is located in the SAL (cf., for instance, 
\citealt{TannerBasuDemarque2012}). 
For very deep extrapolations of more than about two pressure scale
heights, $S_{\rm inflow}$ can no longer be considered a constant, since 
it will slightly increase even inside the nearly adiabatic part of the convection 
zone, in which case necessary changes to $S_{\rm inflow}$ would have to be 
estimated from solar (resp.\ stellar) structure models. Otherwise, with this procedure 
a stable initialisation of simulations of deep stellar surface convection such as that 
one of our Sun is readily possible.

\section{Results}

In the following, we show results from a couple of simulations of solar surface convection
with ANTARES. With these simulations, we explore the merits of the methods presented in 
Section~\ref{setup}.

These results can be generalised to other main sequence stars without any fundamental
complications. Nevertheless, care must be taken in choosing the boundary condition 
parameters adequately. Since this paper is focussed on validating the methods from 
Section~\ref{setup}, we restrict ourselves in the following to the solar case as the 
best-known example of stellar surface convection.

\subsection{Top Boundary Conditions}\label{Res-topBC}

There is one free parameter in the top boundary conditions described in Section~\ref{topbc}.
Changing the parameter $c_{\rm f}$ controls the speed at which $\langle \epsilon \rangle_0$, the value of the
specific internal energy in the upper boundary layers, is adjusted towards the values 
inside the domain (cf., eqs.~\eqref{eq-epsilon0} and~\eqref{eqdelta}).

In the solar case, typical values of the mean speed of sound $\langle v_{\rm snd} \rangle_1$ 
and the mean pressure scale height $\langle H \rangle_1$ in the uppermost domain layer 
are $7\,{\rm km}/{\rm s}$ and $100\,{\rm km}$, respectively. Given a time step size per 
Runge--Kutta integration of around $0.06\,{\rm s}$, the value of $10^{-3}$ given for $\delta$ in 
\citet{Cheung2006} corresponds to $c_{\rm f} \approx 4.0$.

\begin{figure}
\includegraphics[width=1.0\columnwidth]{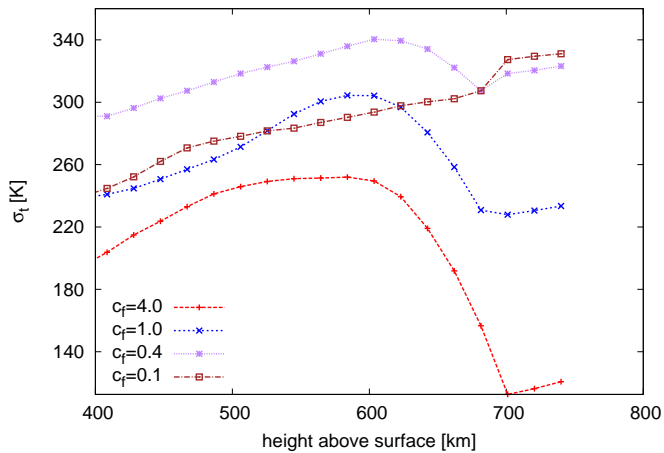}
\caption{Vertical profiles of $\sigma_t$, quantifying the temporal variation of 
         the mean temperature (see \eqref{eq-sigt}), obtained for different values 
         of $c_{\rm f}$ as defined in \eqref{eqdelta}.}
\label{figcf300}
\end{figure}

In two-dimensional numerical experiments we found that the value $c_{\rm f}=4.0$ 
leads to a very slow change in $\langle \epsilon \rangle_0$. We ran a simulation 
with $c_{\rm f}=4.0$ until convection was fully developed. From this point, we started 
the simulations with several values of $c_{\rm f}$ and relaxed them over half an hour. 
Subsequently, they were averaged over another half an hour simulation time.
The (temporal) standard deviation of the mean temperature stratification $\sigma_t$ 
of these models,

\begin{equation}
  \sigma_t \left( T \right) (x) = \langle \left[ \langle T \rangle(x) 
              - \langle\negthinspace\langle T \rangle\negthinspace\rangle_t(x) \right]^2 
              \rangle_t, \label{eq-sigt}
\end{equation}

\noindent where $\langle T \rangle$ is the horizontal and $\langle\negthinspace\langle T 
\rangle\negthinspace\rangle_t$ the temporal and horizontal average of the temperature 
$T$, is plotted in Figure~\ref{figcf300}. 

The rightmost three grid points in this figure show the boundary layers. A value 
of $c_{\rm f}=4.0$ decreases the variation to an undesirably small level at the top, 
which eventually leads to an unrealistic bump in the average $T(\tau)$ relation for 
layers near the top boundary due to a forced slow relaxation, whereas $c_{\rm f}=0.1$ 
leads to much stronger variations in the same region. We suggest to use a value of 
$c_{\rm f} \approx 0.4$ or less. Overall, the effect of $c_{\rm f}$ on the mean 
temperature stratification throughout the model domain is quite small and 
restricted to the upper layers of the simulation starting around $250\,{\rm km}$ 
above the surface. Nevertheless, the choice of large $c_{\rm f}$ can hinder a fast 
relaxation to a different statistically stationary temperature structure. On the other 
hand, very small values of $c_{\rm f}$ may lead to strong variations in the temperature 
which can decrease the numerical stability of the boundary condition. The value of 
$c_{\rm f}=0.4$ suggested here leads to reasonably fast time scales while keeping the 
simulation stable for the solar case. We do not expect this choice of $c_{\rm f} = 0.4$ 
to be universal, and recommend to reassess it for other stellar parameters.

\subsection{Bottom Boundary Conditions}\label{Res-bottomBC}

We compare three 3D solar models with the same simulation setup and different
boundary conditions. The simulation boxes extend to $5.2\,{\rm Mm}$ in the vertical and 
$9.0\,{\rm Mm}$ in the horizontal direction.
In the vertical direction, the resolution is $15.3\,{\rm km}$ whereas in the horizontal 
directions, we chose a resolution of $32.1\,{\rm km}$. The stellar surface is at a 
geometrical depth of around $0.6\,{\rm Mm}$ to $0.7\,{\rm Mm}$ from the top of the 
simulation box. For the radiative transfer equation, we use the grey approximation in 
the upper 30 \% of the box, corresponding to a geometrical depth of around $1.6\,{\rm Mm}$, 
and the diffusion approximation elsewhere. 

Model~2 using the \texttt{BC~2} as in \citet{VoeglerShelyagSchuessleretal2005} was 
initialised with a one-dimensional profile and an initial perturbation. After two hours
of simulation time, the total kinetic energy reached a quasi-constant state, and we 
considered the model to be relaxed. Models~3a and~3b using the boundary conditions 
\texttt{BC~3a} and \texttt{BC~3b}, respectively, were started from later snapshots of
Model~2. After two additional convective turnover times, we started the production runs. 
All production runs cover one hour. In Table~\ref{tab-botBC}, the boundary conditions 
used for these models are summarised.

\begin{table*}[ht]
\begin{minipage}{1.0\textwidth}
\begin{tabular}{l|lllll}
name           & type   & detailed description                     
                        & equations 
                        & time scales $\left[ {\rm h} \right]$        
                        & parameters                   \\
\hline
\texttt{BC~1}  & closed & \citet{MuthsamKupkaLoew-Basellietal2010} 
                        & \eqref{eq-closed}            \\
\texttt{BC~2}  & open   & \citet{VoeglerShelyagSchuessleretal2005} 
                        & \eqref{velbot}, \eqref{epsFrad} \& \eqref{eq-muram} 
                        & $\tau_{\rm KH}\approx  550$
                        &                              \\
\texttt{BC~3a} & open   & \multirow{2}{*}{similar to \citet{FreytagSteffenLudwigetal2012}}
                                                                   
                        & \eqref{velbot}, \eqref{SFtot}, \eqref{Seq1} \& \eqref{eq-co5bold} 
                        & $\tau_{\rm S} \approx  1000$
                        & $\delta_p=1.0$        \\
\texttt{BC~3b} & open   &                                          
                        & \eqref{velbot}, \eqref{SFtot}, \eqref{Seq1} \& \eqref{eq-co5bold}
                        & $\tau_{\rm S} \approx 100$
                        & $\delta_p=0.1$
\end{tabular}
\caption{Summary of bottom boundary conditions in ANTARES. In the fourth column, 
         we refer to the equations constituting these boundary conditions. The 
         time scales are approximate values since they are set in units of sound
         crossing times in the program.} 
        \label{tab-botBC}
\end{minipage}
\end{table*}

\subsubsection{Time Scales}

\begin{figure}
\includegraphics[width=1.0\columnwidth]{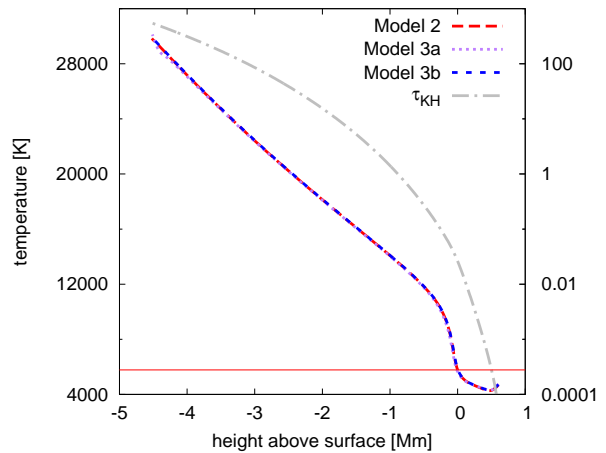}
\caption{Temperature stratification of of Models~2, 3a and 3b, and 
         $\tau_{\rm KH}$ as a function of depth. The effect of the boundary conditions 
         on the mean stratification is rather small, at least after one hour of 
         simulation time. The red horizontal line indicates the effective temperature.}
\label{figTemp}
\end{figure}

The radiative flux at the top of the domain depends on the amount of energy flowing 
through the simulation domain. The idea of formula~(20) in 
\citet{VoeglerShelyagSchuessleretal2005}, i.e.\ eq.~\eqref{epsFrad} in this paper, 
therefore is to correct the (unknown) entropy of the inflow at the bottom boundary 
according to the radiative flux at the top of the domain until the radiative flux has 
reached the desired value. The correction works on the Kelvin--Helmholtz time scale 
$\tau_{\rm KH}$. $\tau_{\rm KH}$ increases rapidly with box depth, as shown in 
Figure~\ref{figTemp}. For our standard simulations of solar surface granulation with a 
domain reaching three to five ${\rm Mm}$ below the stellar surface, $\tau_{\rm KH} 
\approx 100\,{\rm h}$ to $600\,{\rm h}$. In contrast, the time scale for the surface 
granulation is several minutes.

\begin{figure}
\includegraphics[width=1.0\columnwidth]{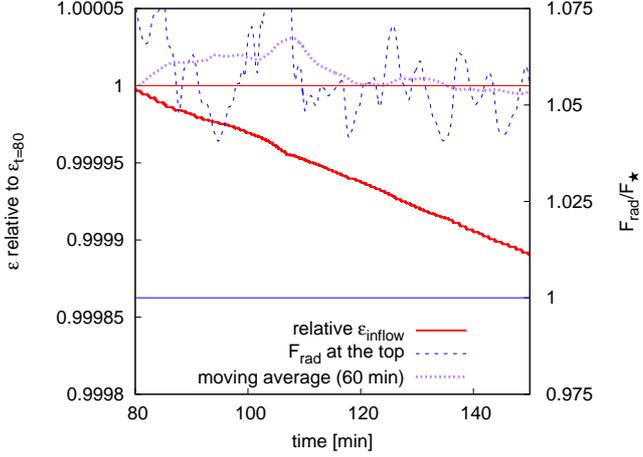}
\caption{Effect of calculating $\epsilon_{\rm inflow}$ by~\eqref{epsFrad} as in 
         Model~2 (red line). The radiative flux $F_{\rm rad}^{\rm top}$ is calculated as 
         horizontal average of $F_{\rm rad}$ in layer $1$ and normalised by the nominal 
         (desired) flux $F_{\star} = \sigma T_{\rm eff}^4$. The red horizontal line 
         indicates the initial value of $\epsilon_{\rm inflow}$, and the blue horizontal 
         line the nominal flux $F_{\star}$. $\tau_{\rm KH} = 600\,{\rm h}$ for this model.}
         \label{figepsFrad}
\end{figure}

\begin{figure}
\includegraphics[width=1.0\columnwidth]{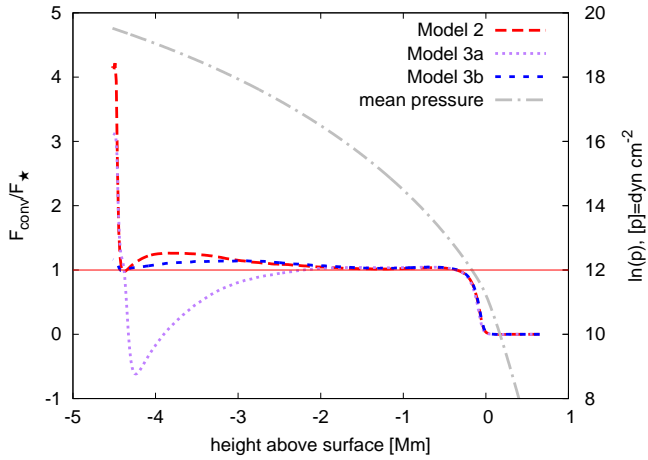}
\caption{Convective flux of Models~2, 3a and 3b normalised by $F_{\star} = \sigma 
         T_{\rm eff}^4$. The mean pressure profile from Model~2 is shown, too (the 
         pressure profiles of the models do not differ significantly). The boundary 
         conditions affect the lower $2\,{\rm Mm}$, or two pressure scale heights, of 
         the domain. All time averages cover one hour of simulation time.}
\label{figFconv}
\end{figure}

\begin{figure}
\includegraphics[width=1.0\columnwidth]{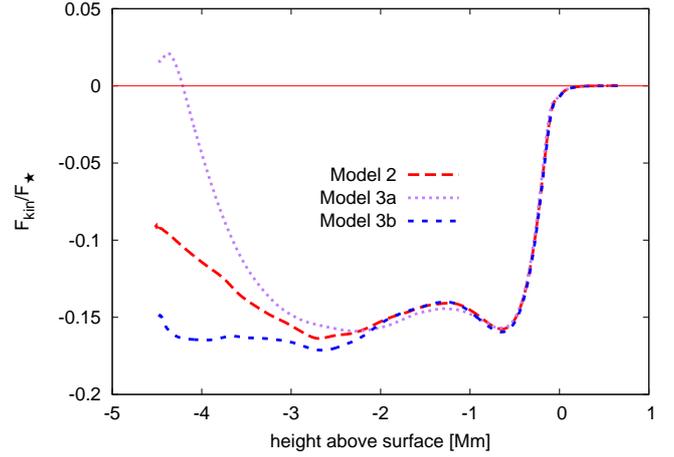}
\caption{Kinetic flux of Models~2, 3a and 3b normalised by $F_{\star} = \sigma 
         T_{\rm eff}^4$. For Model~3a, the kinetic flux changes sign indicating
         that the region covered by downflows is larger than the one covered by 
         upflows.}
\label{figFkin}
\end{figure}

In Figure~\ref{figepsFrad}, the radiative flux at the top of the domain of Model~2 
and $\epsilon_{\rm inflow}$ calculated with~\eqref{epsFrad} are plotted. It is obvious 
from Figure~\ref{figepsFrad} that the correction at the bottom boundary does not influence
the radiative flux at the top of the domain during the time span covered by the 
simulation. But the increasingly strong correction leads to exorbitant changes and 
pathological energy fluxes, as shown in Figs.~\ref{figFconv} and~\ref{figFkin}.
We note that the time span covered by the simulation is less than $1\,\%$ of
$\tau_{\rm KH}$.

$\tau_{\rm KH}$, the time scale of the correction, is around $600\,{\rm h}$ in 
Model~2. We can expect that after this time, the correction at the lower boundary will 
affect the radiative flux at the top of the domain. But performing simulations for such
long times is not feasible with today's computational resources. Common time scales in 
simulations of solar and stellar surface granulation are much shorter. The formulae 
\eqref{epsFrad} and~\eqref{SFrad} require a much longer relaxation time of the model to 
be effective. Therefore, we refrain from using this kind of corrections.

We note here that the simulations presented in \citet{VoeglerShelyagSchuessleretal2005} 
have much more shallow domains with a vertical extent of typically just $1.4\,{\rm Mm}$ 
of which only $0.8\,{\rm Mm}$ are located below a mean depth for which the optical depth 
is $1$. As a result, in their case $\tau_{\rm KH}\approx 1$ to $2\,{\rm h}$. For this 
special case we can expect a much tighter coupling of bottom boundary and 
$F_{\rm rad}^{\rm top}$ than we observe here for our much deeper models because the 
distance between the bottom boundary and the stellar surface is only about two pressure 
scale heights in which case a correction procedure as given by \eqref{epsFrad} can still 
be effective.

Instead, we decided to set the inflowing entropy by formula~\eqref{SFtot} such 
that total flux at the lower boundary $F_{\rm tot}^{\rm bot}$ equals the nominal 
(desired) flux $F_{\star} = \sigma T_{\rm eff}^4$. In Models~3a and~3b, we use the 
boundary conditions \texttt{BC~3a} and \texttt{BC~3b} as specified in 
Table~\ref{tab-botBC} which differ in their choice of $\tau_{\rm S}$ and 
$\delta_p$.

\begin{figure}
\includegraphics[width=1.0\columnwidth]{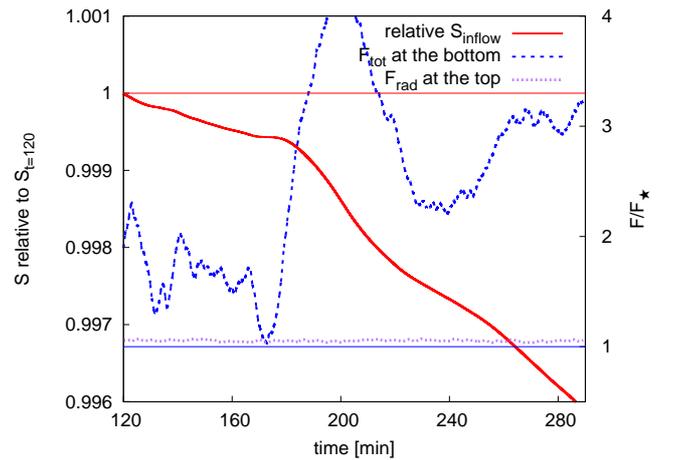}
\caption{Effect of calculating $S_{\rm inflow}$ with~\eqref{SFtot} and
         $\tau_{\rm S} = 1000\,{\rm h}$ as in Model~3a (red line). The radiative flux 
         $F_{\rm rad}^{\rm top}$ is calculated as horizontal average of $F_{\rm rad}$ in 
         layer $1$. $F_{\rm rad}$ and $F_{\rm tot}^{\rm bot}$ are normalised by 
         $F_{\star} = \sigma T_{\rm eff}^4$. The red horizontal line indicates the 
         initial value of $S_{\rm inflow}$, and the blue horizontal line the nominal 
         flux $F_{\star}$.}
         \label{figSFtot3a}
\end{figure}

\begin{figure}
\includegraphics[width=1.0\columnwidth]{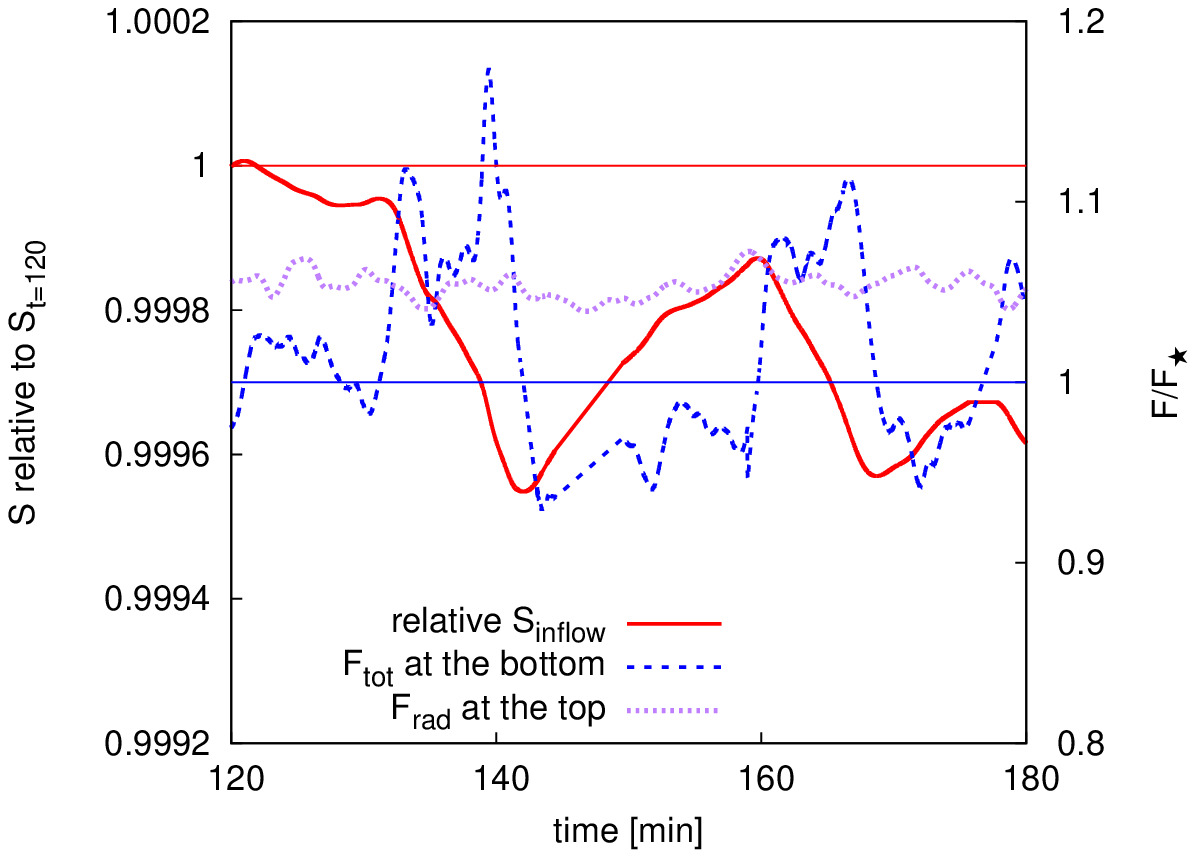}
\caption{Effect of calculating $S_{\rm inflow}$ with~\eqref{SFtot} and
         $\tau_{\rm S} = 100\,{\rm h}$ as in Model~3b (red line). The radiative flux 
         $F_{\rm rad}^{\rm top}$ is calculated as horizontal average of $F_{\rm rad}$ in 
         layer $1$. $F_{\rm rad}$ and $F_{\rm tot}^{\rm bot}$ are normalised by 
         $F_{\star} = \sigma T_{\rm eff}^4$. The red horizontal line indicates the 
         initial value of $S_{\rm inflow}$, and the blue horizontal line the nominal 
         flux $F_{\star}$.}
         \label{figSFtot3b}
\end{figure}

$\tau_{\rm S}$ must be set according to the time scale on which $F_{\rm tot}^{\rm bot}$
changes. As can be seen in Figure~\ref{figSFtot3a}, the value of $1000\,{\rm h}$ chosen
for $\tau_{\rm S}$ in Model~3a is much too large. The change in entropy is too slow to 
keep $F_{\rm tot}^{\rm bot}$ at the desired level. The resulting convective and kinetic 
fluxes are pathological, as can be seen in Figs.~\ref{figFconv} and~\ref{figFkin}.
On the other hand, the correction works perfectly with $\tau_{\rm S} = 100\,{\rm h}$ as
shown in Figure~\ref{figSFtot3b} for Model~3b. The overall changes in entropy 
and energy fluxes are small, and the resulting mean fluxes are reasonable.

In Figure~\ref{figTemp}, we show the temperature profile of these models. The influence 
of the drastic changes in the boundary conditions and in the energy fluxes is hardly 
visible on this time scale, but will increase with time. In any case, we prefer 
\texttt{BC~3b} as used in Model~3b, since these boundary conditions keep the energy 
fluxes in the lower part of the domain at the desired level and do not produce any 
serious artifacts even if the initial value of $S_{\rm inflow}$ is not chosen very well.

\subsubsection{Dependence on Parameters of the Boundary Conditions}

The parameter $\delta_p$ in \citet{FreytagSteffenLudwigetal2012} defines a time 
scale on which pressure fluctuations at the bottom boundary are damped. $\tau_{\rm S}$ 
controls the time scale on which $S_{\rm inflow}$ is changed. Models~3a and~3b 
differ only in the values chosen for these parameters: for Model~3a, we set 
$\delta_p=1.0$ and $\tau_{\rm S} \approx 1000\,{\rm h}$. For Model~3b, 
$\delta_p=0.1$ and $\tau_{\rm S} \approx 100\,{\rm h}$. Both use the 
formula~\eqref{SFtot} for the calculation of $S_{\rm inflow}$. Model~2 uses 
\texttt{BC~2}. 

\begin{figure}
\includegraphics[width=1.0\columnwidth]{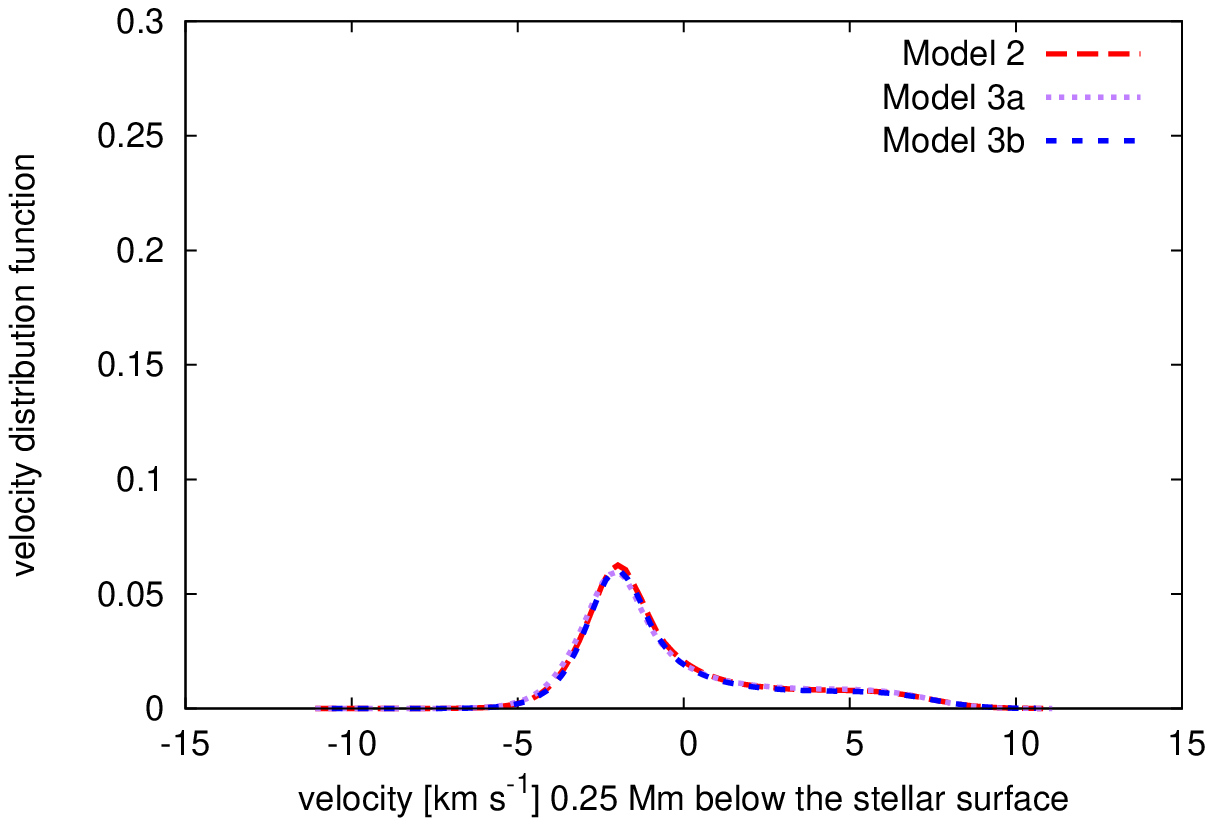}
\hfill
\includegraphics[width=1.0\columnwidth]{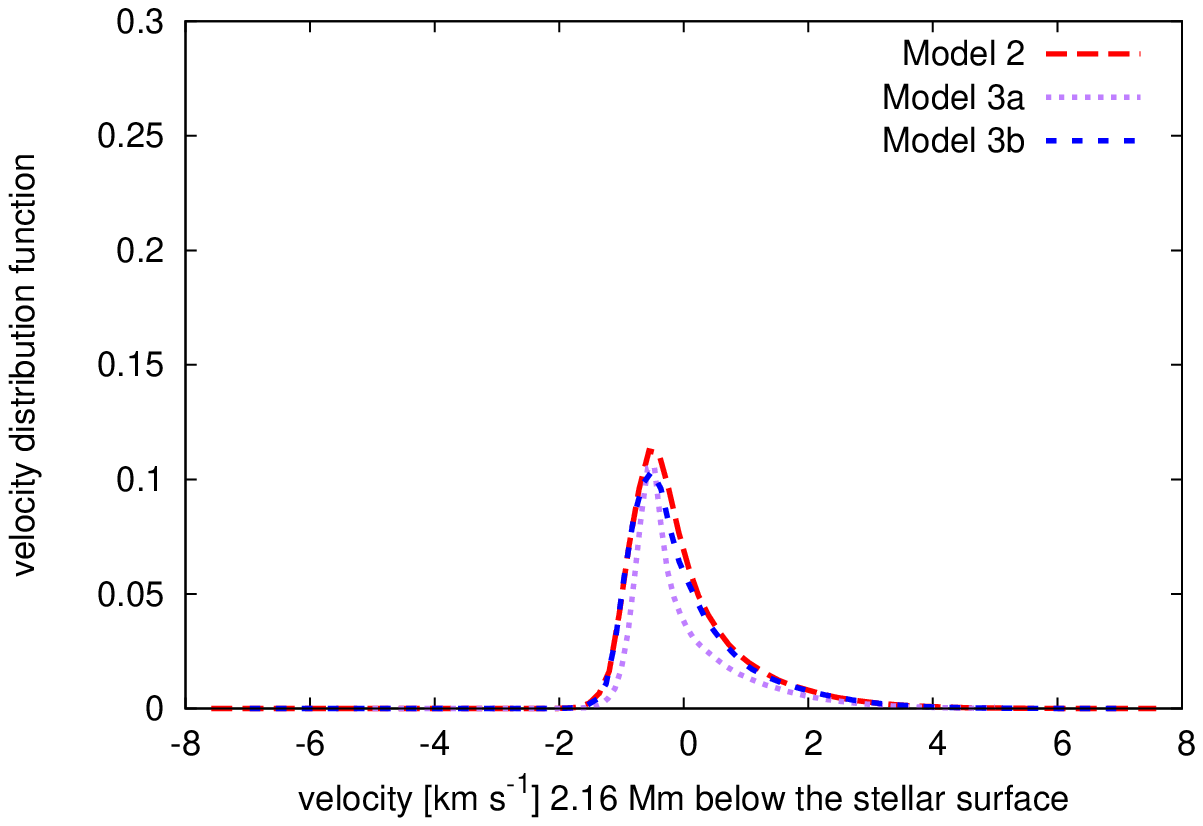}
\hfill
\includegraphics[width=1.0\columnwidth]{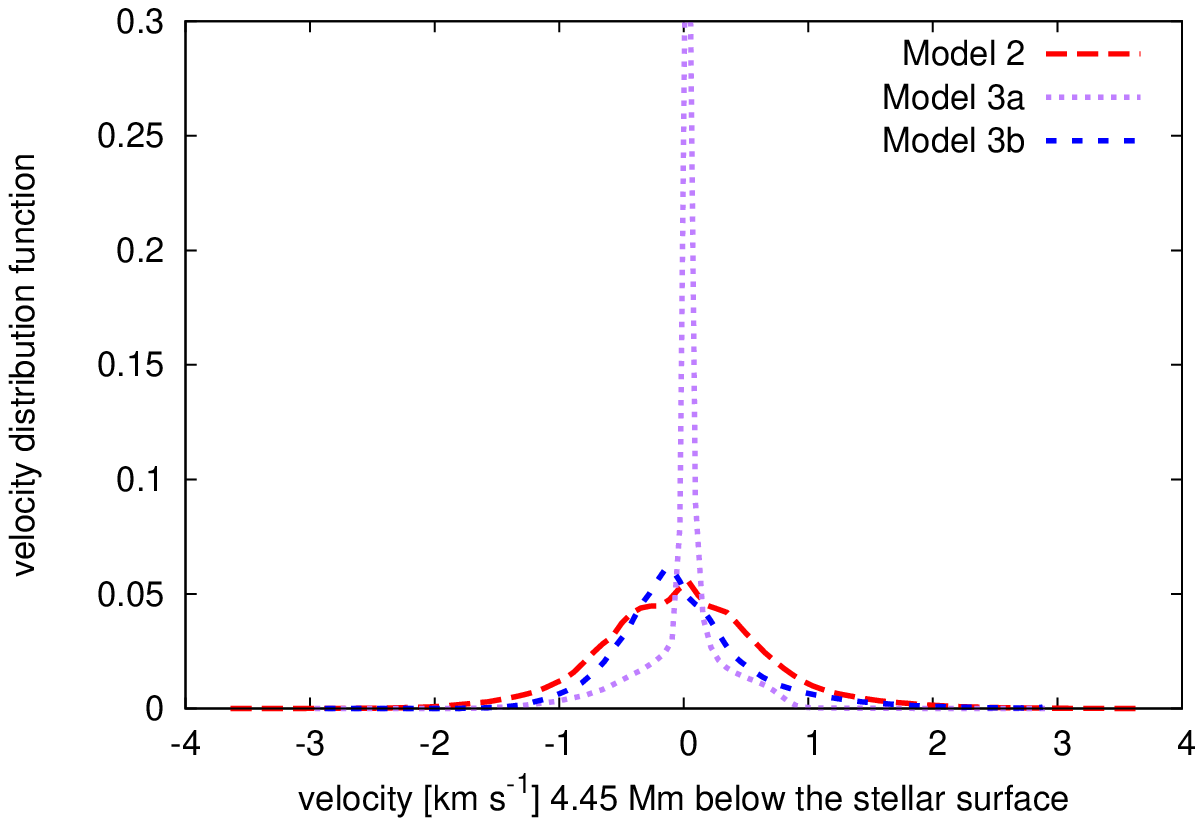}
\caption{Velocity distribution $0.25\,{\rm Mm}$, $2.16\,{\rm Mm}$ and $4.45\,{\rm Mm}$ 
         below the stellar surface (from top to bottom). The lowermost layer 
         is only $0.1\,{\rm Mm}$ above the lower boundary and therefore strongly 
         influenced by the boundary conditions. For these pictures, the 
         velocities in each of the layers were grouped into $96$ equal-sized bins. 
         The range of the bins depends on the maximum vertical velocity in this layer.  
         Then, the distribution function was normalised by the number of nodes 
         and time steps such that the total sum is $1$. Due to the 
         coordinate system chosen for ANTARES upflows imply $u(x,y,z) < 0$.}
         \label{scvel}
\end{figure}

In Figure~\ref{scvel}, we show the distribution of the vertical velocity of the three 
models at three depths. At the depth shown in the top panel of Figure~\ref{scvel}, which 
is situated $0.25\,{\rm Mm}$ below the stellar surface, the differences between the 
three models are very small. As expected, the distribution has a significant 
skewness due to the asymmetry between the faster, narrow and turbulent downflows and 
the slower, more uniform upflows. Note the shift of the maximum induced by the broad 
upflows (and the fully compressible treatment of the flow). Going deeper and therefore 
closer to the bottom boundary the differences between the models become more prominent. 
In the bottom panel, the distribution of Model~3a has much narrower tails and its 
skewness is much smaller than in Model~2 or Model~3b, whereas the velocity distribution 
of Model~2 and Model~3b is in all three layers very similar (as we expect it to be). 
On the other hand, there are some deviations visible for Model~2 while for Model~3b, 
the distribution is almost invariant as a function of depth apart from a 
gradual broadening observed for layers farther away from the stellar surface. By 
modifying only two parameters in the bottom boundary conditions, we changed the 
velocity distribution of Model~3a in the lower third of the simulation domain 
considerably: the distribution approaches a narrow Gaussian one with a very flat, 
additional tail and, in the end, resulting in a small and positive kinetic energy flux.

Comparing our results with the data from \citet{SteinNordlundGeorgovianietal2009}, 
who performed simulations of the upper part of the solar convection zone in boxes 
of up to $20\,{\rm Mm}$ depth, we conclude that there is no physical reason for 
the kinetic flux to go to $0$. On the contrary, it should even slightly increase 
in magnitude when going deeper down into the convection zone --- even though 
that trend in the kinetic flux in the lower part of their simulation box may also be
influenced by their boundary conditions.

Furthermore, we conclude from Figure~3 in \citet{SteinNordlundGeorgovianietal2009}, 
that the asymmetry between up- and downflows persists when going deeper into the 
convection zone, while simultaneously the dominant horizontal scale of the 
convection increases. This means that the velocity distribution function with depth 
should have similar shapes.

Returning to the differences between Model~2 and Model~3b, in the bottom panel 
of Figure~\ref{scvel}, the distribution of velocities in Model~2 is closer to 
symmetric at the bottom, whereas for Model~3b, the asymmetry is less affected by 
the boundary. The same can also be observed when comparing more shallow 
models such as those summarised in Table~\ref{tabFrad}. What we observe for 
Model~3b is hence in agreement with \citet{SteinNordlundGeorgovianietal2009}, 
which is not the case for Models~2 and~3a as their behaviour as a function of
depth may be modified by just changing the depth of the simulation box. Therefore 
boundary condition \texttt{BC~3b} is the best choice among the three models we 
investigated.

Individual testing of the parameters $\tau_{\rm S}$ and $\delta_p$ would 
have not been affordable in terms of computation time. Anyway, the combination of 
$\tau_{\rm S} \approx 100\,{\rm h}$ and $\delta_p=0.1$ as in \texttt{BC~3b} 
yields sensible results and we choose them as new default values for these parameters.

\subsection{Dependence on Initial Model}\label{initialmodel}

In the following, we compare three solar 3D models which differ only in their 
initial conditions, three different 1D solar structure models. {\it model\,S} 
from \citet{Christensen-Dalsgaardetal1996} combines an MLT model of convection 
\citep{Boehm-Vitense1958} with $\alpha_{\rm MLT}=1.99$ used to compute the 
solar envelope with a semi-empirical model for the photosphere. This model
has been calibrated by its authors to match the observed solar radius and luminosity 
of the Sun at its present age. Moreover, it reproduces the depth of the solar convection
zone as inferred from helioseismology \citep{RosenthalChristensen-DalsgaardNordlundSteinTrampedach1999}. 
It is our standard initial model and was used for all simulations of solar surface convection 
in this paper unless explicitly mentioned.
The second and third one are purely theoretical models. They are calculated with 
the CESAM code \citep{Morel1997,MorelLebreton2008} and use the grey Eddington approximation 
for the photosphere. The implementation of the convection model of CESAM is described in
\citet{SamadiKupkaGoupiletal2006}. In the second model, convection is modelled by means of 
an MLT model with $\alpha_{\rm MLT}=1.766$. The third one uses the CGM model of convection 
\citep{CanutoGoldmanMazzitelli1996} with $\alpha_{\rm CGM} = 0.69$. All differences between 
the models are summarised in Table~\ref{tabinitmodels}.

Since the 1D structure models do not reach high enough into the atmosphere, we extended 
the 1D models to cover the $4\,{\rm Mm}$ vertical extent of our simulations with 
$\sim 700\,{\rm km}$ above the stellar surface. All our 3D simulations treat radiative 
transfer with the opacity binning method with $4$ bins and were relaxed for $80$ to 
$90\,{\rm min}$ with a vertical resolution of $20\,{\rm km}$. Then, the resolution 
was increased to about $11\,{\rm km}$ in the vertical and $35\,{\rm km}$ in the horizontal 
directions. After two sound crossing times, we started the production run with a duration 
of $30\,{\rm min}$.
All 3D simulations use the open boundary conditions \texttt{BC~3b} at the bottom,
the LLNL equation of state \citep{RogersSwensonIglesias1996}, the non-grey opacities 
from \citet{Kurucz1993CD13,Kurucz1993CD2} and the opacity data from \citet{IglesiasRogers1996} 
for the deep interior. For the composition of \citet{GrevesseNoels1993}, we confirmed 
a good agreement between our choices of atmospheric and interior opacities in the transition 
region where $\tau \sim 1000$ and $T \sim 10{,}000\,{\rm K}$.

\begin{table*}[ht]
\begin{minipage}{1.0\textwidth}
\begin{tabular}{l|llll|lll|ll}
model 
      & atmosphere
      & convection
      & $\alpha$
      & $S_{\rm bot}$
      & $X$ & $Y$ & $Z$  
      & $L_x$ & $L_P$ \\
 & & & & $10^9 \cdot \left[{\rm erg}\,{\rm g}^{-1}\,{\rm K}^{-1} \right]$
 & & & & $\left[{\rm Mm}\right]$ & $\left[{\rm Mm}\right]$ \\
\hline
model\,S & semi-empirical & MLT & $1.99$ 
         & $1.77355$ & $0.7373$ & $0.2427$ & $0.0200$ 
         & $4.07$ & $0.75$ \\
CGM      & Eddington & CGM &  $0.69$ 
         & $1.77760$ & $0.7381$ & $0.2438$ & $0.0181$ 
         & $4.09$ & $0.77$ \\
MLT      & Eddington & MLT & $1.766$ 
         & $1.77680$ & $0.7381$ & $0.2438$ & $0.0181$
         & $4.04$ & $0.72$
\end{tabular}
\caption{Differences in convection and atmosphere model as well as composition 
         for the three models from Section~\ref{initialmodel}. We also give the
         vertical box length $L_x$ as well as $L_P$, the height of the region where 
         $\langle T \rangle < T_{\rm eff}$. At the beginning of the simulation, 
         $S_{\rm inflow}$ is initialised by the entropy of the initial model at the 
         position of the lower boundary denoted by $S_{\rm bot}$ in this table.}
         \label{tabinitmodels}
\end{minipage}
\end{table*}

\begin{figure*}
\includegraphics[width=1.0\columnwidth]{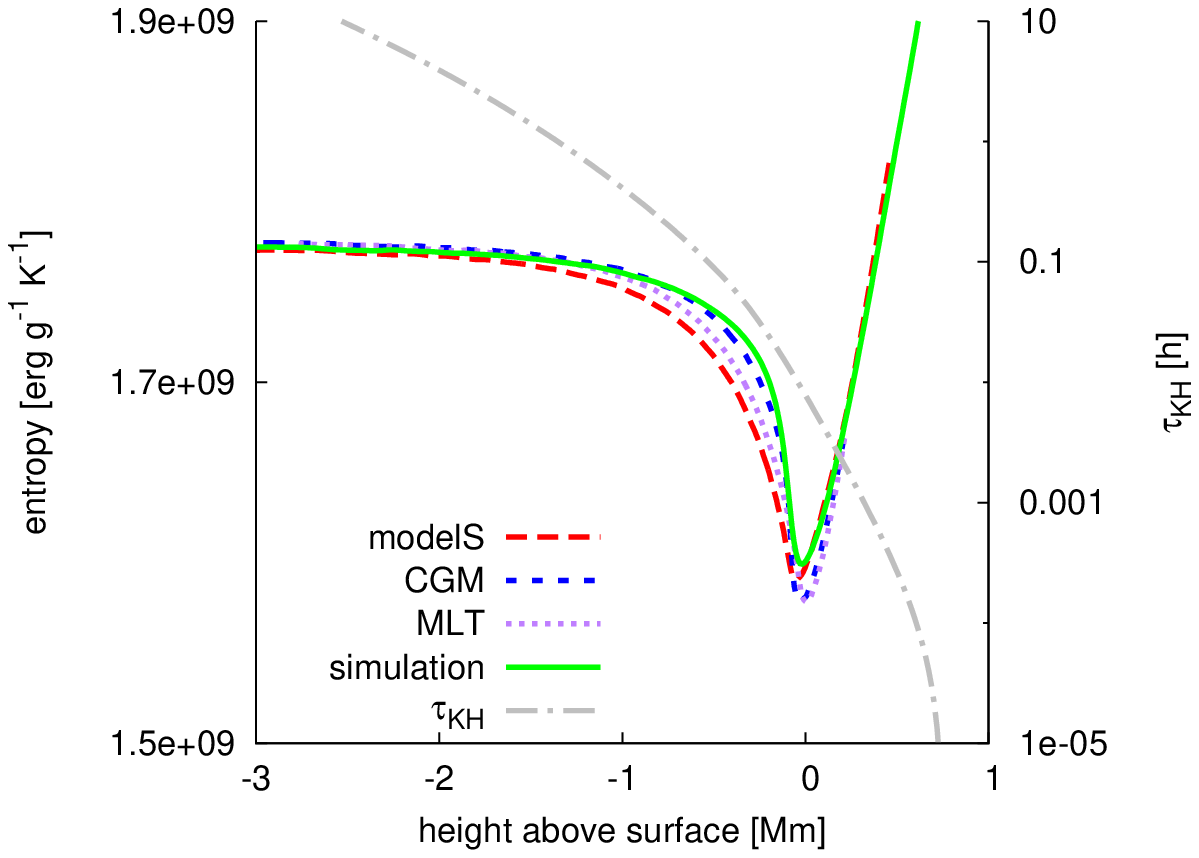}
\includegraphics[width=1.0\columnwidth]{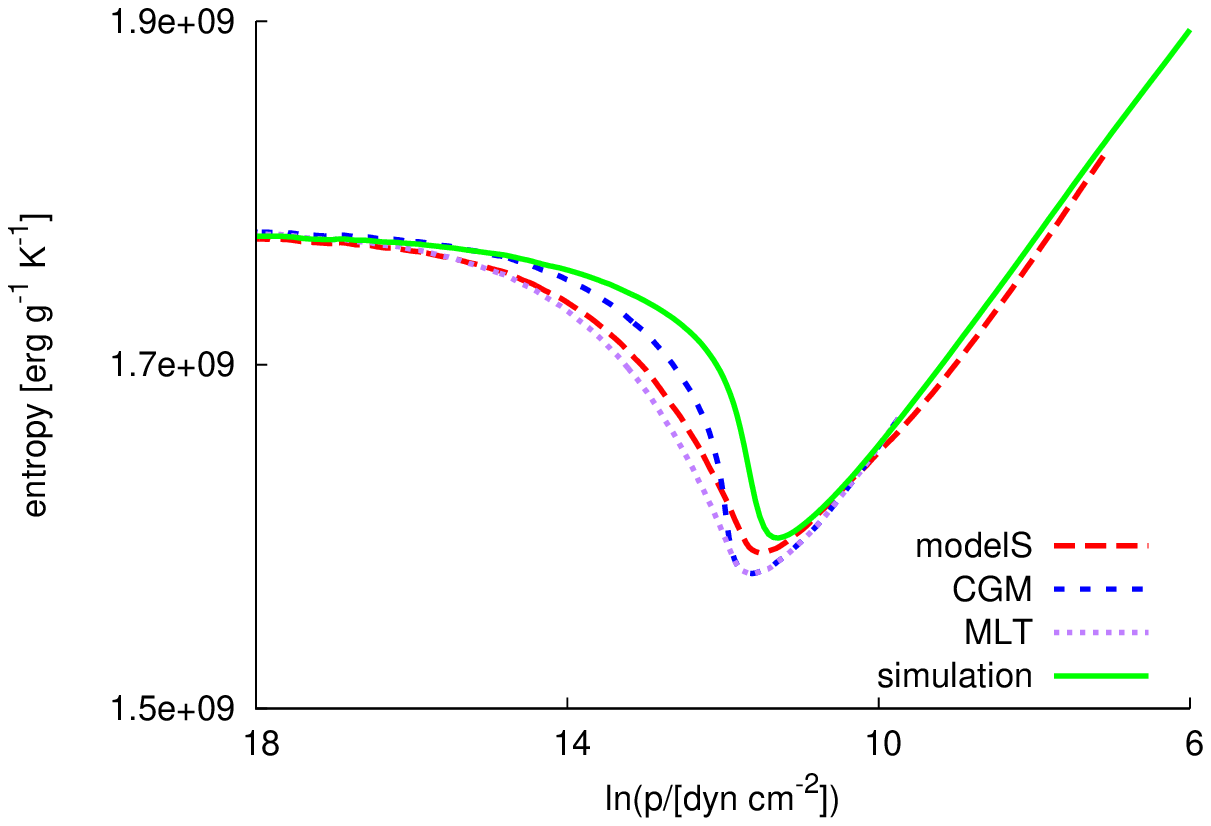}
\caption{Entropy of the three initial models ({\it model\,S}, CGM, MLT) and of the relaxed
         numerical simulation as well as $\tau_{\rm KH}$ as a function of depth (left)
         and of the natural logarithm of pressure (right). Since the simulations based 
         on these initial conditions all feature similar relaxed profiles, only one 
         representative simulation is shown in its relaxed state, too. On the left,
         all curves are aligned such that the region where $\langle T \rangle = T_{\rm eff}$
         coincides.}
         \label{figentropy}
\end{figure*}

As depicted in Figure~\ref{figentropy}, the initial entropy profiles differ mainly 
in the layers from $1\,{\rm Mm}$ below the surface up to the surface itself. At the 
bottom boundary, the differences in entropy are much smaller. Immediately below the 
superadiabatic peak, the three profiles from the relaxed simulations, however, are 
very similar to the CGM initial model. As they do not differ considerably from each 
other, only one simulation profile is shown in Figure~\ref{figentropy}. Since 
$\tau_{\rm KH}$ is about $1$ to $2\,{\rm h}$ in this region, the relaxation time was 
long enough to allow the simulations to adjust to a common stratification. Near the 
surface, the simulation results differ from all three one-dimensional models.

\begin{figure*}
\includegraphics[width=1.0\columnwidth]{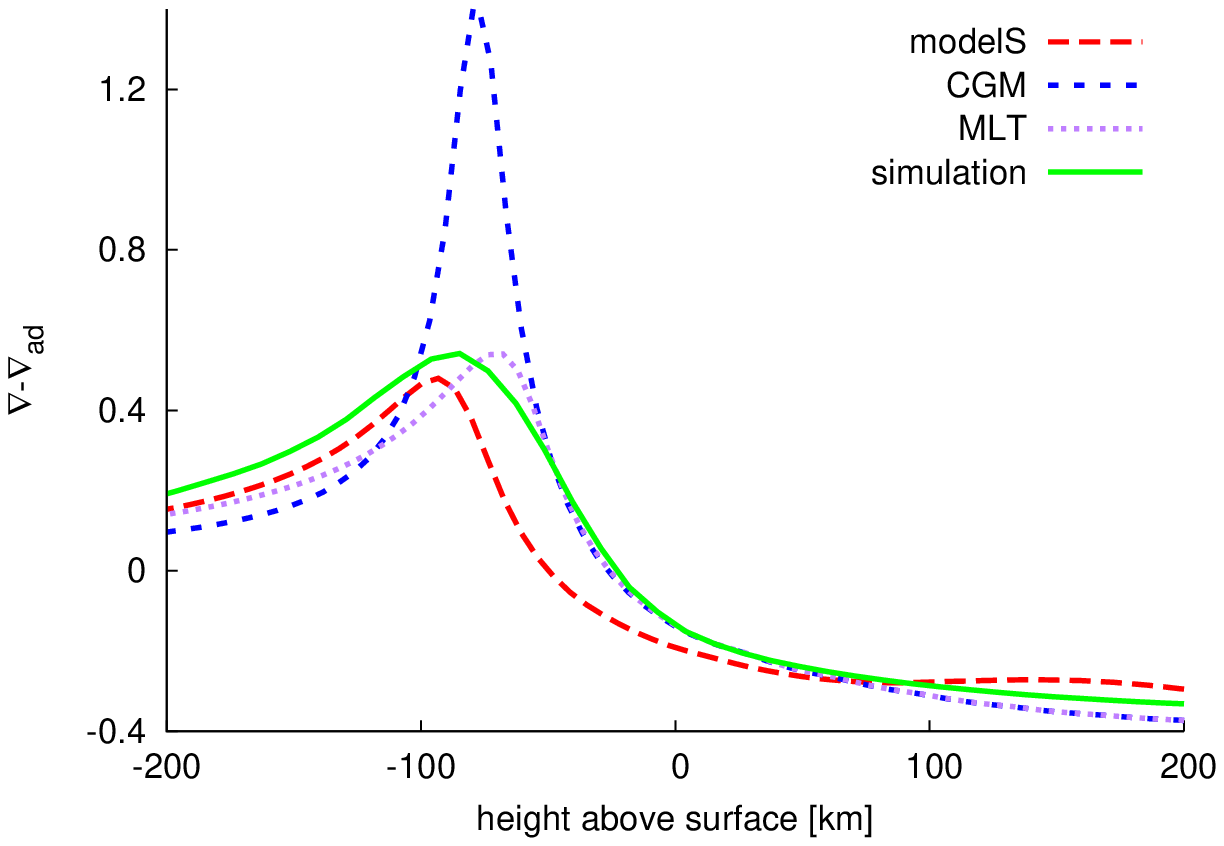}
\includegraphics[width=1.0\columnwidth]{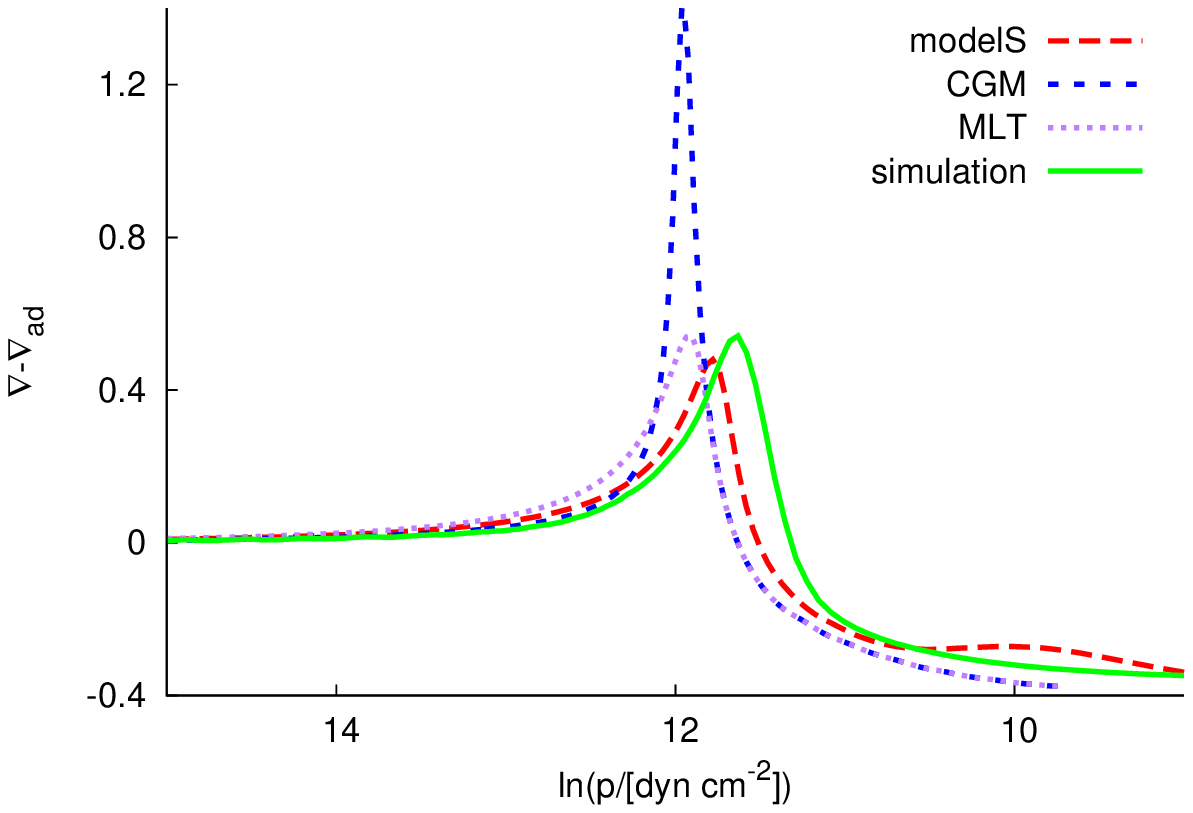}
\caption{$\nabla - \nabla_{\rm ad}$ of the three initial models and of the numerical 
         simulation as discussed in Fig.~\ref{figentropy}. The integration rule from 
         \citet{Carlson1962} was used for the angular integration in the radiative 
         transfer solver. The logarithmic temperature gradient $\nabla$ is computed 
         with respect to the total (i.e.\ including turbulent) pressure. On the left,
         all curves are aligned such that the region where $\langle T \rangle = T_{\rm eff}$
         coincides.}
\label{fignabla}
\end{figure*}

In the right panel of Figure~\ref{figentropy} where the mean entropy profile is 
plotted over mean logarithmic pressure, we clearly observe the elevation of the optical 
surface in the fully developed three dimensional simulation \citep{NordlundStein1999}.
The simulation profile is shifted by about one pressure scale height compared to the 
MLT and {\it model\,S} profile. Figure~\ref{fignabla} demonstrates how this leads to a similar
shift upwards for the peak of superadiabicity $\nabla - \nabla_{\rm ad}$ (cf.\ 
\citealt{KimChan1998}). $\nabla - \nabla_{\rm ad}$ in the SAL as shown in 
Figure~\ref{fignabla} is very steep for the CGM initial model. Therefore, the CGM model 
is able to yield an entropy profile close to the one from the simulations in the interior, 
while the profile of the simulation above the SAL is closer to that of the MLT model.

In all three models, once relaxed, the entropy profile as well as the radiative flux at 
the top do not differ significantly from each other. Thus, differences in the SAL exhibited 
by the initial models are not important, if the simulations are run long enough in terms of 
$\tau_{\rm KH}$ for that region. From Figure~\ref{figentropy} we deduce that $\tau_{\rm KH}$ 
is about $1$ to $2\,{\rm h}$ in the region where the entropy profiles of the initial models 
differ, which corresponds to the simulation time used for relaxation. We return to the 
role of $\tau_{\rm KH}$ in Section~\ref{section-discussion}.

\subsection{Resolution Dependence}\label{Res-resolution}

Since {\it model\,S} provides the correct adiabat for the solar case, we can 
expect our simulations to reproduce $T_{\rm eff}$ of the Sun. Therefore, we can 
use the criterion $T_{\rm eff} = T_{\rm eff,\odot}$ as necessary condition for 
the correctness of our models.
From numerical experiments we deduce that the vertical resolution directly influences 
$T_{\rm eff}$ of the numerical model. A numerical resolution of at least $10$ to 
$11\,{\rm km}$ is necessary for this purpose, if the integration rule of 
\citet{Carlson1962} is used for the angular integration. These numbers are slightly 
different depending on the type of binning used for the radiative transport, and 
depend also on the quadrature formula used to perform angular integration.

The resolution dependence is demonstrated in Table~\ref{tabFrad}, where a few 
grey and non-grey models are compared with respect to the radiative flux at the top 
of the domain, including Models~2, 3a and~3b from Section~\ref{Res-bottomBC}. 
The other models differ compared to them in terms of simulation domain size, 
duration of the simulation and grid resolution. Grid resolution, box size, the 
geometrical depth where the stellar surface is situated, and the bottom boundary 
conditions are specified in Table~\ref{tabFrad}.
All models are three-dimensional and the radiative transfer equation is solved along 
$24$ rays using the quadrature rule of \citet{Carlson1962} for the angular integration, 
but the number of bins differs. The grey approximation corresponds to the case of one 
bin, whereas the non-grey models were calculated with four bins. Details can be 
found in~\ref{app-fluxes} and \citet{MuthsamKupkaLoew-Basellietal2010}.

In the grey case, $F_{\rm rad}^{\rm top}$ is slightly lower than in the non-grey 
case where we used $4$ bins (cf.\ models with a vertical resolution of $11.0\,{\rm km}$ 
and $13 .0\,{\rm km}$, respectively, corresponding to a difference in $T_{\rm eff}$ 
between $48\,{\rm K}$ and $55\,{\rm K}$). 
$F_{\rm rad}^{\rm top}$ strongly depends on the vertical resolution. 
We emphasize that the rate of change of $F_{\rm rad}^{\rm top}$ with numerical 
resolution depends on the angular quadrature rule. The rule by \citet{Carlson1962}, 
which does not include a vertical ray, requires a very high resolution to yield the 
correct radiative flux (only the non-grey model with $6.5\,{\rm km}$ vertical 
resolution yields a $T_{\rm eff}$ very close to the solar one).

Conversely, the dependence of $F_{\rm rad}^{\rm top}$ on the lower boundary 
conditions is weak on the time scales of these simulations. Table~\ref{tabFrad} 
does not show any systematic difference with respect to the bottom boundary conditions. 
Changes due to boundary effects occur only on much longer time scales.

We note that the horizontal resolution of all simulations presented here is sufficient 
to get the correct effective temperature and radiative flux at the top of the domain 
(cf.\ \citealt{AsplundLudwigNordlundStein2000,RobinsonDemarqueLietal2003}).

\begin{table*}[t]
\begin{center}
\begin{tabular}{llllll|l|l|cc}
opacities & $\Delta x$ & $\Delta y$ 
          & $L_x$      & $L_y$      & $L_P$
          & b.\ c.\    & $T_{\rm eff}$ 
          & $S_{\rm bot}$ & $S_{\rm bot} - S_{\rm min}$ \\
          & \multicolumn{2}{c}{$\left[ {\rm km} \right]$}
          & \multicolumn{3}{c|}{$\left[ {\rm Mm} \right]$} 
          & & $\left[ {\rm K} \right]$ 
          & \multicolumn{2}{c}{$10^9 \cdot \left[{\rm erg}\,{\rm g}^{-1}\,{\rm K}^{-1} \right]$} \\
\hline
grey     & $19.5$ & $40.0$ & $4.0$ & $6.0$ & $0.63$ & \texttt{BC~2}  & $5959.8$ & $1.77131$ & $0.16418$ \\ 
grey     & $13.0$ & $40.0$ & $4.0$ & $6.0$ & $0.65$ & \texttt{BC~2}  & $5835.8$ & $1.77179$ & $0.17576$ \\ 
grey     & $13.0$ & $40.0$ & $4.0$ & $6.0$ & $0.65$ & \texttt{BC~3b} & $5834.7$ & $1.77167$ & $0.17546$ \\ 
grey     & $11.0$ & $35.3$ & $4.1$ & $6.0$ & $0.74$ & \texttt{BC~3b} & $5813.7$ & $1.77219$ & $0.17800$ \\ 
non-grey & $13.0$ & $40.0$ & $4.0$ & $6.0$ & $0.66$ & \texttt{BC~3b} & $5884.2$ & $1.77095$ & $0.17296$ \\ 
non-grey & $11.0$ & $35.3$ & $4.0$ & $6.0$ & $0.66$ & \texttt{BC~3b} & $5868.1$ & $1.77065$ & $0.17325$ \\ 
non-grey & $ 6.5$ & $20.0$ & $4.0$ & $6.0$ & $0.68$ & \texttt{BC~2}  & $5765.3$ & $1.77015$ & $0.18086$ \\
\hline
Model~2  & $15.3$ & $32.1$ & $5.2$ & $9.0$ & $0.65$ & \texttt{BC~2}  & $5850.2$ & $1.77203$ & $0.17379$ \\
Model~3a & $15.3$ & $32.1$ & $5.2$ & $9.0$ & $0.65$ & \texttt{BC~3a} & $5853.5$ & $1.77252$ & $0.17409$ \\
Model~3b & $15.3$ & $32.1$ & $5.2$ & $9.0$ & $0.65$ & \texttt{BC~3b} & $5853.9$ & $1.77277$ & $0.17451$ \\
\end{tabular}
\caption{Effective temperature calculated from $F_{\rm rad}^{\rm top} = 
         \sigma T_{\rm eff}^4$ in dependence of binning method, numerical resolution and 
         boundary conditions. $\Delta x$ is the grid spacing and $L_x$ the box length
         in the $x$ (vertical) direction, $\Delta y$ and $L_y$ in the $y$ direction. Note that 
         $\Delta z = \Delta y$ and $L_z = L_y$ in all cases. $L_P$ is the height of the region 
         where $\langle T \rangle < T_{\rm eff}$.
         In addition, the entropy at the bottom of the computational domain and the entropy 
         jump is given. The observed effective temperature of the Sun is $5777.6\,{\rm K}$. 
         Boundary conditions are explained in Table~\ref{tab-botBC}. We also show the values 
         for the three models from Section~\ref{Res-bottomBC} which use the grey approximation 
         for radiative transfer.}
         \label{tabFrad}
\end{center}
\end{table*}

\subsection{Comparison with Closed Boundary Conditions}\label{Res-openclosed}

\begin{figure}
\includegraphics[width=1.0\columnwidth]{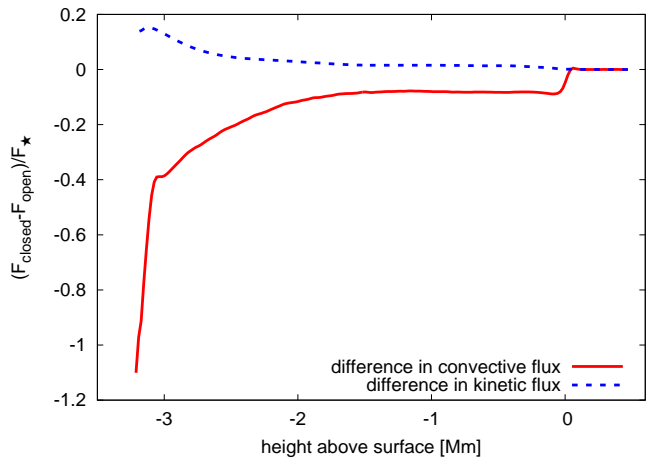}
\caption{Difference in the energy fluxes of a model with closed and a model with 
open boundary conditions normalised by $F_{\star} = \sigma T_{\rm eff}^4$.}
\label{figdifffluxes}
\end{figure}

In Figure~\ref{figdifffluxes}, the energy fluxes of two similar models, one 
with closed boundaries \texttt{BC~1} and one with open boundaries \texttt{BC~3b},
are compared. The time average extends over two hours for the model with closed 
boundaries, and over one hour for the model with open ones. Since the resolution 
of the model with open boundary conditions was higher ($11.0\,{\rm km}$ vs.\ 
$19.4\,{\rm km}$ in the vertical direction), the data was interpolated 
to the coarser grid of the model with closed boundaries. The resolution changes 
the radiative flux near the surface, similarly to what was shown in 
Table~\ref{tabFrad}, thus explaining the step in $F_{\rm conv}$ near a height of 
$0\,{\rm km}$. In the following, we only discuss the differences in the kinetic 
and convective fluxes.

The closed boundary conditions force a zero convective and zero kinetic energy flux
at the boundaries. Compared to the model with open boundary conditions the convective 
and the kinetic flux stay smaller in magnitude --- keeping in mind their opposite sign. 
This can be explained by considering the fact that the numerical resolution of 
the model with closed boundary conditions is rather coarse leading to a wrong radiative 
flux, and that the top boundary is quite close to the stellar surface.

Forcing the vertical velocities to $0$ at the bottom boundary influences the energy 
fluxes on about two pressure scale heights. The effect on the kinetic energy flux is smaller 
than on the convective flux. Around $1\,{\rm Mm}$ above the bottom boundary, the kinetic 
energy flux from the model with closed boundaries reaches a saturated state with very small 
differences to the model with open boundaries.

\subsection{Two--Dimensional Models}\label{Res-twodim}

As demonstrated in Figures~\ref{figepsFrad}, \ref{figSFtot3a}, \ref{figSFtot3b} 
and~\ref{figentropy}, thermal relaxation of a simulation takes place on 
Kelvin--­Helmholtz time--scale of the simulation \citep{Kupka2009b}. Much care should 
therefore been taken in setting up the initial state of a simulation, which should be 
close to the relaxed, statistical steady state. This is most important for the deepest 
parts of the simulation where the Kelvin­--Helmholtz time scale is long. Thus, if a good 
initial model is not available, a three-dimensional simulation may not be run long 
enough with reasonable effort such that it can thermally relax properly.

Since the computational costs for a two-dimensional model are much lower, one is 
tempted to relax the model in two dimensions first and switch to three dimensions 
as soon as the correct amount of energy is contained in the simulation domain. 
Unfortunately, we have found that two- and three-dimensional models 
of solar surface convection behave quite differently during relaxation.

\begin{figure}
\includegraphics[width=1.0\columnwidth]{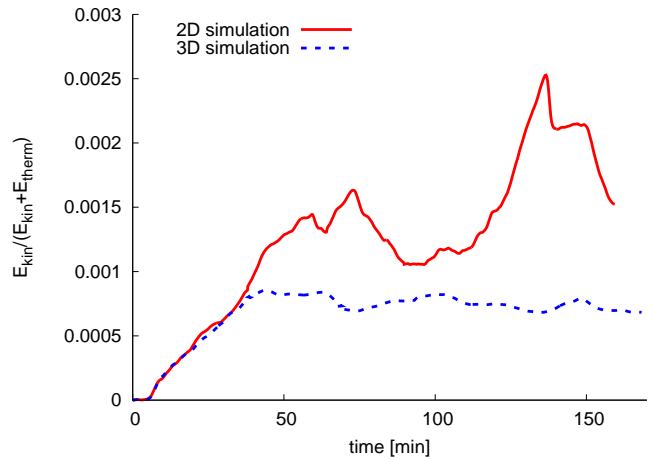}
\caption{Temporal evolution of the normalised total kinetic energy for 
         a two- and a three-dimensional simulation.}
\label{figtekina}
\end{figure}

In Figure~\ref{figtekina}, the time evolution of the total kinetic energy $E_{\rm kin}$ 
scaled by the sum of kinetic and thermal energy of a two- and a three-dimensional 
simulation both starting from the same one-dimensional model is shown. 
In theory, the kinetic energy should saturate as soon as the convection is fully 
developed, and the simulation should reach a quasi-stationary state. Obviously, 
this happens quite rapidly in the three-dimensional case, whereas the kinetic 
energy of the two-dimensional model starts to oscillate with a very high amplitude 
and without showing any signs of saturation.

From Figure~\ref{figtekina} we deduce that a two-dimensional simulation intended 
as a precursor of a three-dimensional simulation should be conducted only for a 
few first sound crossing times. Even then, it can be hard to get rid of the artifacts 
of the two-dimensionality, especially if the model reaches deep into the convection 
zone. When we tried to convert a two-dimensional model which was run for a long time 
(around $60$ sound crossing times) to a three-dimensional one, we encountered 
serious problems destroying the symmetry in the third dimension in the deeper 
layers, and the relaxation to a truly three-dimensional state was not shorter than 
when the model were started directly from the one-dimensional initial model. On the 
contrary, the model inherits a lot of the deficiencies of the two-dimensional model 
such as statistical non-stationarity and an overall higher vorticity.

\citet{AsplundLudwigNordlundStein2000} pointed out that a two-dimensional 
simulation of solar granulation leads to different stratifications than a 
three-dimensional simulation and that the results from two-dimensional
simulations are qualitatively wrong concerning line profiles and element 
abundances. Furthermore, they emphasized that two- and three-dimensional 
simulations need different values for the entropy of the inflowing material 
(about $2 \%$ difference) to reach the desired effective temperature. 
As a consequence, it appeared advisable to us to avoid using two-dimensional 
simulations for testing the bottom boundary conditions since the resulting 
entropy profiles will differ from their three-dimensional counterparts.

Nevertheless, pre-computing a model in two dimensions can be useful if the 
two-dimensional phase does not take too long, and if the model is shallow 
such that the time scale of motion at the bottom boundary is still quite 
short. For instance for the solar case, the first hour after starting from 
the one-dimensional initial model may be calculated in two dimensions, if 
the model is not deeper than $4\,{\rm Mm}$.

Figure~\ref{figtekina} furthermore demonstrates that the thermal 
stratification of a model is quite robust against kinetic motions 
since the kinetic energy reaches only about $0.1 \%$ of the thermal 
energy. This ratio is comparable to the ratio of $\tau_{\rm sound}$ 
to $\tau_{\rm KH}$ and thus essentially the time scale of convective
transport relative to the time scale of simulation relaxation from our
``arbitrary'' initial conditions (see Sect.~\ref{section-discussion}).

Confirming the results from Section \ref{initialmodel} convective 
motions would need a long time to change the thermal 
stratification of the entire model, especially since quasi-adiabaticity 
keeps the sizes of the fluxes of kinetic energy and enthalpy within the same 
order of magnitude. A significantly different stratification could 
of course lead to a higher total flux which in turn could reduce the 
relaxation time somewhat. Still, a good starting model is an essential 
prerequisite for cost efficient numerical simulations.

\begin{figure}
\includegraphics[width=1.0\columnwidth]{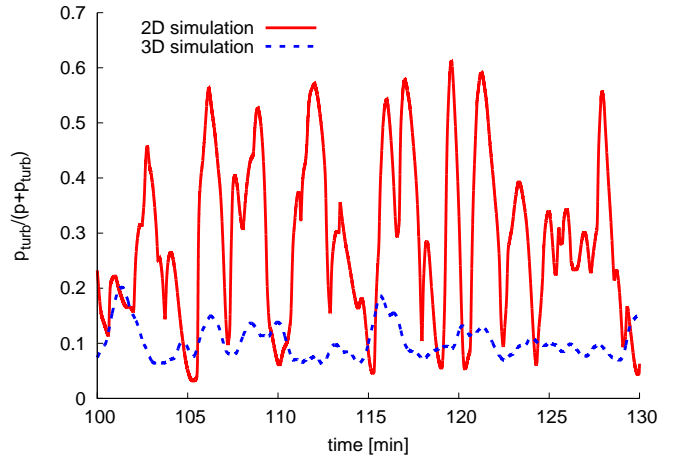}
\caption{Temporal evolution of the ratio $p_{\rm turb} / (p+p_{\rm turb})$ 
         in the innermost boundary layer for the same simulations.}
\label{figPturb}
\end{figure}

In Figure~\ref{figPturb} another difference between two- and 
three-dimensional simulations is shown. There, the relative share 
of the horizontally averaged turbulent pressure in the innermost 
boundary layer of the open top boundary condition is plotted. The 
simulations are the same as in Figure~\ref{figtekina}. 
In both cases, $c_{\rm f}=4.0$, but we tested several different 
values and found no noticeable variation. The difference
in the relative contribution of turbulent to total pressure is 
significant: in the three-dimensional case, it fluctuates around 
$10 \%$ whereas in two dimensions, the turbulent pressure sometimes 
even exceeds the sum of the gas and the radiative pressure. In the 
mean, its share is around $30 \%$ demonstrating once more the unstable 
and non-stationary nature of the two-dimensional simulation for the 
upper layers of the (solar) photosphere.

Still, in both cases, the share of the turbulent pressure in the boundary 
layers is not negligible. Using \eqref{modhydstat} is therefore 
definitely an improvement compared to using the hydrostatic equation 
without turbulent pressure. Nevertheless, Figure~\ref{figPturb}
demonstrates that any static approximation of these layers is rather 
unrealistic.

We note that in comparison with Figure~1.3 (bottom) in \citet{Steffen2000}, 
the difference between two and three dimensions is of a similar magnitude, 
even though the layer depicted in Figure~\ref{figPturb} is even higher 
in the photosphere than the case represented by the data from 
\citet{Steffen2000}.

\section{Discussion}\label{section-discussion}

The physical realism of simulations of stellar surface convection depends strongly 
on the boundary conditions imposed at the top and the bottom of the simulation 
domain. In a large part of the simulation domain, the flow is influenced in a 
direct way by the boundary conditions. Therefore, the simulation box must be wide 
enough in all directions, and the boundary should disturb the velocity field
as little as possible. 
With closed boundary conditions, the motion of the fluid is forced to stop at the 
boundaries, influencing the dynamics in a large part of the simulation domain 
(cf.\ also \citealt{KupkaRobinson2007,Kupka2009b}). Open boundary conditions are clearly 
preferable as they allow free motion of the fluid, but even with open boundary 
conditions we recommend to ensure a distance of at least two pressure scale heights 
between the lower boundary and the SAL.

At the top of the domain, time scales are very short so that effects of the initial 
stratification are rather short-lived. The effects of the upper boundary are limited 
by the amount of mass contained there (compared to the rest of the simulation), so 
locating the top boundary sufficiently far above the optical surface is crucial.
The boundary conditions~\eqref{eq-topbc} assume an isoenergetic
and hydrostatic stratification, including the effect of turbulent pressure. Since we 
impose constant velocities with depth in the top boundary layers, waves are not reflected,
but transmitted, which increases the long-term stability of the model. The formulation
of the top boundary conditions allows variations in the temperature away from the initial 
profile. In our implementation, the parameter $c_{\rm f}$ controls the stiffness of 
the open boundary conditions in terms of temporal variations of the temperature, 
as shown in Figure~\ref{figcf300}. We found the optimal value by experimentation and 
it will vary with atmospheric parameters.

The most difficult part in designing open boundary conditions at the bottom of the 
simulation domain is to specify the internal energy of the inflows. This value 
controls the overall stratification of a surface convection model by setting 
the adiabat of the deep (isentropic) convection zone. But the coupling 
with the energy radiated at the top of the domain operates on time scales which are,  
for deep, nearly adiabatic convection zones, not feasible to be calculated with 
multi-dimensional simulations.

For radiative (and conductive) zones a derivation of why $\tau_{\rm KH}$ can be used as 
an approximation to the {\em thermal adjustment time}, which is the actual quantity of 
interest here, can be found in Chapter~5.3 of \citet{KippenhahnWeigert1994}, where also 
limitations of this approximation are pointed out. For the case of convection zones the 
approximation $\tau_{\rm KH} \approx \tau_{\rm adj}$ is derived in Chapter~6.4 of the same 
reference as part of a general stability analysis for stellar material at a certain (arbitrary) 
depth. But for the models considered in Section~\ref{Res-bottomBC}, $\tau_{\rm KH}$
is about $6000$ times larger than the convective turn-over time, i.e.\ the total time 
for gas to rise from the bottom to the surface, and go back out the bottom boundary of 
the simulation again. Therefore, a thermal relaxation of the models on $\tau_{\rm KH}$ 
as intended by formulae~\eqref{epsFrad} and~\eqref{SFrad} is not affordable with the 
available computation power.
As illustrated with Figure~\ref{figentropy}, differences in the entropy of the initial model 
within the superadiabatic region disappear within the local $\tau_{\rm KH}$ whereas the 
stratification of simulations over shorter relaxation times is determined by the initial 
model. The success of one-dimensional models in predicting the nearly adiabatic 
stratification for most but about the upper $1\,{\rm Mm}$ of the solar convection zone 
is essential for the efficiency of hydrodynamical simulations and for the short relaxation 
times of 1 to 2 hours of numerical simulation of solar granulation reported in the literature.

For simulations of solar convection, any correction mechanism which correlates the bottom 
boundary directly with the top boundary will be inefficient, if the simulation box spans 
more than $1$ or $2\,{\rm Mm}$ of depth below the stellar surface. On the other hand, in shallow 
simulation boxes the boundary conditions will directly influence the SAL. 
The simulation box must be large to keep the influence of the bottom boundary small. 
This, however, prohibits a direct correction mechanism of the entropy at the bottom boundary 
based on the radiative flux on the top. Instead, the input entropy (resp.\ internal 
energy) is much better controlled by a local criterion like~\eqref{SFtot}. We prefer 
to impose the boundary conditions on the entropy since its value is nearly constant with depth,
such that the boundary conditions can be applied at a range of geometrical depths using the same 
parameter values. Criteria like~\eqref{epsFrad} or~\eqref{SFrad} can lead to wrong energy 
fluxes as shown in Figures~\ref{figFconv} and~\ref{figFkin}, since the correction at the 
bottom does not affect the upper boundary in the time covered by the simulation. In terms 
of Table~\ref{tab-botBC}, our preferred boundary conditions are \texttt{BC~3b}. These 
boundary conditions allow free in- and outflow with an adiabatic inflow, while fixing 
the mean mass flux over the bottom boundary to $0$. We refrain from using \texttt{BC~2} 
and \texttt{BC~3a} for future simulations.

In the formulation of open boundary conditions, several parameters must be set which 
influence the behaviour of the flow near the physical boundaries considerably. But, since
the computational costs for each single simulation of surface convection are high,
systematic tests of the specific effect of each parameter are expensive. Nevertheless,
we found a reliable choice for all parameters at least for the solar case, which should
be widely applicable to stars of the central and lower main-sequence.

The computational costs cannot be reduced by performing simulations only in two 
dimensions, since the results differ considerably. As stated also in 
\citet{AsplundLudwigNordlundStein2000}, two-di\-men\-sional simulations lead to 
a different stratification and convective velocities since the inherent 
convective transport properties are different. Furthermore, longer simulation 
times are needed for statistical significance, since the area coverage of 
two-dimensional simulations is smaller --- which negates the original gain in 
computational speed. We have found that in addition two-dimensional simulations of 
solar surface convection do not reach an energetic equilibrium on the same time scale 
as three-dimensional ones and the upper photosphere is influenced by much stronger 
fluctuations of turbulent pressure. As the symmetries in the flow-field are also 
difficult to break by applying a perturbation to initialise a three-dimensional 
simulation, the use of two-dimensional models as precursors of three-dimensional 
ones should hence be limited to only five to ten sound crossing times such
that the two-dimensional simulation has not yet developed its peculiar 
thermal stratification and velocity profile.
Alternatively, horizontal averages may be used to construct one-dimensional
initial models, if the latter cannot be constructed in a reliable way otherwise.

The numerical resolution influences directly the radiative flux at the top of the domain.
As we show by numerical experiments in Table~\ref{tabFrad}, the accuracy of the radiative 
transfer solver can be improved by increasing the numerical resolution on the hydrodynamical 
grid. If the resolution is sufficiently high, the radiative flux and therefore the effective 
temperature is in agreement with the observed values. When the angular quadrature rule 
by \citet{Carlson1962} is used, a vertical grid spacing of at least $10$ to 
$11\,{\rm km}$ or, preferably, even better than $7\,{\rm km}$, is necessary for the solar 
case, otherwise the effective temperature of the model will be too high. Therefore, 
a realistic simulation of surface granulation must be done in sufficiently high 
resolution. If the aspect ratio of each computational cell is very different from $1$, 
the magnitude of the intrinsic viscosity of the numerical scheme varies in dependence 
of the coordinate direction, and surfaces which are inclined with respect to the coordinate 
system are not well resolved. Therefore, increasing the vertical resolution must be 
accompanied by an increase in horizontal resolution as well, which makes the simulation 
much more expensive.

The required spatial resolution depends on the angular integration rule used 
in the radiative transfer solver. \citet{TannerBasuDemarque2012} discussed that 
the specific intensity in a fixed grid point is highly variable in time and in the 
inclination angle. Therefore, choosing a suitable integration rule for the angular 
dependence of the intensity is crucial to accurately determine the radiative flux, 
and different formulae will give different results. Only in the limit of more and 
more ray directions these results will converge.

In contrast to the angular integration rule from \citet{TannerBasuDemarque2012}, but also
\citet{SteinNordlund2003}, the integration rule used in our simulations \citep{Carlson1962} 
does not have a vertical ray. But at least in the centre of a granule, the radiative flux is 
dominated by the vertical direction. If only few rays are used in the angular integration, as
it is common in stellar surface convection simulations, a vertical ray should be included 
in the integration scheme. We will show in a forthcoming paper, how choosing an 
angular integration rule changes the radiative flux of
a numerical simulation of stellar surface convection.

\section{Conclusions}\label{section-conclusions}

In this paper, we investigate the shortcomings of several formulations of open
boundary conditions and discuss how to implement boundary conditions for a 
high-order method which allow realistic simulations of stellar surface granulation.

At the top of the domain, open boundary conditions can transmit waves originating at 
the stellar surface whereas closed boundaries reflect them. At the bottom boundary, 
the value of the inflowing entropy must be specified. Some of the proposed approaches, 
which couple the input flux at the bottom boundary to $F_{\rm rad}$ at the top of the 
domain \citep{VoeglerShelyagSchuessleretal2005}, work only for shallow boxes where the 
Kelvin--Helmholtz time scale is small. Also, $F_{\rm rad}$ at the surface of a model is 
sensitive to the numerical resolution. The initial model determines the stratification 
of the model, which can only change on the Kelvin--Helmholtz time scale.
Open bottom boundary conditions can be expected to excel over closed ones only, if they 
work on the right time scales, are combined with a suitable, local criterion for the 
inflow entropy, and if the simulation domain is sufficiently large in all directions.
Keeping the inflow entropy variable in time allows control of the total energy 
flux at the bottom boundary.

The high computational requirements of each single simulation of stellar surface 
convection prohibits systematic testing of each parameter in the formulation of 
the open boundary conditions.
Nevertheless, we have found a reliable choice for both top and bottom boundaries, i.e.\  
\texttt{BC~3b} from Table~\ref{tab-botBC}, which has become the default for numerical 
simulations of stellar surface convection with ANTARES.

\section*{Acknowledgements}

We acknowledge financial support from the Austrian Science fund (FWF), projects 
P20973 and P20762. FK acknowledges support by the FWF grant P21742. 
HGS wants to thank H.~Muthsam for helpful discussions and the MPA Garching for 
a grant for a research stay in Garching. 
Calculations have been performed at the VSC clusters of the Vienna universities 
and the Heraklit cluster of the TU~Cottbus.
The model with closed boundary conditions has been calculated at RZG.
We would like to thank G.~Houdek for making {\it model\,S} and R.~Samadi for making 
the MLT and the CGM initial models available to us.
We are thankful to K.~Belkacem and J.~Ballot for carefully reading the manuscript 
and suggesting a number of improvements. We thank the referees for their helpful 
comments.




\appendix

\section{Damping Initial Oscillations}
\label{app-damp}

Since multidimensional convection models set up from one-dimensional stellar 
models are not in their lowest energy state, they start to oscillate in the 
vertical direction \citep{TrampedachAsplundColletNordlundStein2013}.
These artifical oscillations differ in amplitude and frequency from the p--mode 
oscillations expected to be seen in the simulation of stellar surface convection.
To remove them, we adapted the procedure described in detail in \citet{Trampedach1997}
and \citet{TrampedachAsplundColletNordlundStein2013} in our code. The damping term is 
an additional term in the velocity equation. The damping velocity $v_{\rm mode}$ 
is defined by

\begin{equation}
  v_{\rm mode} = \frac{\langle\rho u\rangle}{\langle\rho\rangle}.
\end{equation}

The (vertical) momentum equation takes the form

\begin{equation}
  \frac{\partial \rho u}{\partial t} = {\rm  rhs\ terms\ from\ eq.~\eqref{eq-NS}} 
                                     - \rho\,\frac{v_{\rm mode}}{t_{\rm mode}},
\end{equation}

\noindent where $t_{\rm mode}$ is the period of the mode which should be 
damped. To be consistent, we also added a corresponding term to the total energy 
equation such that

\begin{equation}
  \frac{\partial E}{\partial t} = {\rm  rhs\ terms\ from\ eq.~\eqref{eq-NS}} 
                                - \rho u\,\frac{v_{\rm mode}}{t_{\rm mode}},
\end{equation}

\noindent which improves the stability of this procedure and restores the conservation
of energy (even though the conservation errors of the original method are very small). 
During a production run, we do not apply any damping of vertical momentum.

This procedure can easily be extended to remove dispensable horizontal momenta by 
replacing the vertical velocity in the calculation of $v_{\rm mode}$ by its 
horizontal counterparts $v_{{\rm mode},y}$ or $v_{{\rm mode},z}$:

\begin{equation}
  v_{{\rm mode},y} = \frac{\langle\rho v\rangle}{\langle\rho\rangle},\ v_{{\rm mode},z} 
  = \frac{\langle\rho w\rangle}{\langle\rho\rangle}.
\end{equation}

The time scale $t_{\rm mode}$ can be used to control the rate at which the 
horizontal momenta are removed. With values of $t_{\rm mode} \approx 1$ sound 
crossing time, we found that removing of horizontal momenta, if desired, works 
very efficiently. Even in production runs, we damp the horizontal momenta 
in the boundary layers to avoid an excessive increase in horizontal momenta.

\section{Asymmetric Stencils at the Domain Boundaries}
\label{app-asymmetric}

As described in \citet[p. 53]{FerzigerPeric2002}, enforcing conditions 
which set derivatives to zero as in various prescriptions for the
in- and outflow at top and bottom boundaries like~\eqref{veltop},
\eqref{velbot}, or~\eqref{eq-muram} should not be done by simply setting $u$ 
to a constant value, but by exact inversion of the stencils used for the 
calculation of the derivatives. Especially, if higher-order methods as, e.g., 
the fourth order method presented in \citet{HappenhoferGrimm-StreleKupkaetal2013}, 
are used, unsuitable procedures can lead to pathological behaviour of the velocity 
field near the boundary.

With a five-point wide stencil, boundary conditions for ANTARES must be set on three 
vertical layers. As before, all physical layers are numbered from $1$ to $n_x$ starting 
at the top of the simulation box. At the top boundary of the computational domain, 
derivatives of a scalar-valued function $\phi$ at the cell centres are calculated by

\begin{subequations}
\begin{align}
  \frac{\partial \phi}{\partial x}(x_{-2}) & 
    = \frac{ - 25 \phi_{-2} + 48 \phi_{-1} - 36 \phi_{0} + 16 \phi_{1} - 3 \phi_{2}}{12 \Delta x}, \\
  \frac{\partial \phi}{\partial x}(x_{-1}) & 
    = \frac{ -  3 \phi_{-2} - 10 \phi_{-1} + 18 \phi_{0} -  6 \phi_{1} +   \phi_{2}}{12 \Delta x},
\end{align}

\noindent following the approach presented in \citet{HappenhoferGrimm-StreleKupkaetal2013}. 
For $\frac{\partial \phi}{\partial x}(x_{0})$, the symmetric stencil can be used, i.e.\ 

\begin{equation}
  \frac{\partial \phi}{\partial x}(x_{0}) = \frac{ \phi_{-2} - 8 \phi_{-1} + 8 \phi_{1} - \phi_{2}}{12 \Delta x},
\end{equation}\label{eq-diffstencils}
\end{subequations}

\noindent where $\phi_i=\phi(x_i)$ and mirrored expressions are valid for the bottom 
boundary. For $\frac{\partial \phi}{\partial x}(x_{\frac{1}{2}})$ and other derivatives 
at the cell boundary, similar expressions can be calculated with the same procedure.

Expansion in Taylor series shows that the error term $\varepsilon$ is given by

\begin{subequations}
\begin{align}
  \varepsilon \left( \frac{\partial \phi}{\partial x}(x_{-2}) \right) &
                              = - \frac{1}{5}  \left( \Delta x \right)^4 \phi^{(5)}(\zeta), \\
  \varepsilon \left( \frac{\partial \phi}{\partial x}(x_{-1}) \right) & 
                              =   \frac{1}{20} \left( \Delta x \right)^4 \phi^{(5)}(\zeta), \\
  \varepsilon \left( \frac{\partial \phi}{\partial x}(x_{0})  \right) & 
                              = - \frac{1}{30} \left( \Delta x \right)^4 \phi^{(5)}(\zeta),
\end{align} \label{eq-bnddrv}
\end{subequations}

\noindent demonstrating the fourth order of the given stencils. The error constant 
increases the more the stencils are asymmetric.

If we try to enforce $\frac{\partial \phi}{\partial x}=0$ by setting 
$\phi_{-2}=\phi_{-1}=\phi_{0}=\phi_{1}$, the derivatives will not be 
zero numerically. Instead, we solve eqs.~\eqref{eq-bnddrv} for 
$\pad[\phi]{x} = 0$, to get

\begin{subequations}
\begin{align}
  \phi_{-2} = & \frac{64}{55} \phi_{1} - \frac{9}{55} \phi_{2}, \\
  \phi_{-1} = & \frac{63}{55} \phi_{1} - \frac{8}{55} \phi_{2}, \\
  \phi_{0}  = & \frac{64}{55} \phi_{1} - \frac{9}{55} \phi_{2}.
\end{align}\label{eq-invstencils}
\end{subequations}

While this procedure immediately applies to bottom boundary conditions as specified
through \eqref{epsFrad}, \eqref{eq_tau_KH}, and \eqref{eq-muram},
for conditions set by~\eqref{SFrad} or \eqref{SFtot}, \eqref{velbot} in conjunction 
with~\eqref{eq-co5bold}, the vertical velocity $u$ in the innermost boundary 
layer $n_x+1$ is already set by the enforcement of mass conservation~\eqref{eq_rho3}. 
Whereas we can use the procedure from above for the horizontal velocities, the boundary 
condition for the vertical velocity must be changed to

\begin{subequations}
\begin{align}
  u_{n_x+1} & \rm{\ set\ to\ enforce\ mass\ conservation}, \\
  u_{n_x+2} = & \frac{279}{197} u_{n_x+1} -  \frac{99}{197} u_{n_x} +  \frac{17}{197} u_{n_x-1}, \\
  u_{n_x+3} = & \frac{252}{197} u_{n_x+1} -  \frac{64}{197} u_{n_x} +  \frac{9}{197}  u_{n_x-1}.
\end{align}
\end{subequations}

\section{Calculation of the Energy Fluxes}
\label{app-fluxes}

The calculation of the energy fluxes follows \citet{Canuto1997b} which describes
the procedure for the case of a perfect gas. We have generalized this to arbitrary 
equations of state in the following. We also define the residual velocities

\begin{equation}
  u'' = u - \frac{\langle\rho u\rangle}{\langle\rho\rangle},\ 
  v'' = v - \frac{\langle\rho v\rangle}{\langle\rho\rangle},\ 
  w'' = w - \frac{\langle\rho w\rangle}{\langle\rho\rangle},
  \label{eq-Uprime}
\end{equation}

\noindent where net mass fluxes have been removed. For fluxes only, we ignore our 
coordinate convention, and have positive fluxes going out through the top of the box.

\subsection{Radiative Flux ($F_{\rm rad}$)}

In regions where the optical depth is $\lesssim 100$, the radiative 
transfer equation is solved with the short characteristics method as described in 
\citet{MuthsamKupkaLoew-Basellietal2010}. The equation is solved by an angular 
integration over a discrete set of rays. The dependence on the frequency is 
considered by repeating the calculation several times for averages of intensity over 
frequency sets, to each of which the weight $\omega_{\rm bin}$ is assigned 
\citep{Nordlund1982,SteinNordlund2003,LudwigJordanSteffen1994}.
In every grid-point of the simulation, the intensity $I({\rm bin},{\rm ray})$ is 
calculated for $N_{\rm rays}$ rays, where $N_{\rm rays}$ depends on the quadrature 
formula chosen for the angular integration, and $N_{\rm bins}$ frequency bins
\citep{LudwigJordanSteffen1994}. For the integration rule from \citet{Carlson1962} 
which is used in the simulations presented in this paper, $N_{\rm rays}$ is 24 in three 
and 12 in two dimensions due to symmetries. In three dimensions, we have three rays 
per octant with two different angles in latitude and three in the azimuthal direction.
$N_{\rm bins}=1$ is the grey approximation, and $N_{\rm bins}=4$ is used for non-grey 
simulations. The opacity data is taken from \citet{IglesiasRogers1996},
\citet{FergusonAlexanderAllardetal2005} and \citet{Kurucz1993CD13,Kurucz1993CD2}. For 
details, we refer to \citet{MuthsamKupkaLoew-Basellietal2010}.

The vertical component of the radiative flux is calculated by

\begin{equation}
  F_{\rm rad} = \frac{4\pi}{N_{\rm rays}} \sum_{\rm rays}
                \sum_{\rm bins} \omega_{\rm bin}\,n_x\,I({\rm bin},{\rm ray}),
  \label{eqFrad}
\end{equation}

\noindent where $n_x$ is the vertical component of the ray direction. In optically 
thick regions, the diffusion approximation is valid and $F_{\rm rad}$ can be 
calculated simply by

\begin{equation}
  F_{\rm rad} = k \, \frac{\partial T}{\partial x}.
\end{equation}

The sign of the heat conduction coefficient, $k$, is positive due to the coordinate 
system chosen in ANTARES.

In our simulations, the transition to the diffusion approximation is done at a 
fixed geometrical depth. We choose this depth such that the transition is in any case 
performed in a region where the optical depth in every grid point is between $100$ 
and $1000$, also in case of optically thin downflows.

\subsection{Convective Flux ($F_{\rm conv}$)}

The vertical component of the convective flux is defined by

\begin{equation}
  F_{\rm conv} = \langle u'' \rho \left( h - \frac{\langle e+p \rangle}{\langle \rho \rangle} \right) \rangle,\label{eqFconv}
\end{equation}

\noindent where $h = \frac{e+p}{\rho}$ is the specific enthalpy and $u''$
is defined in equation~\eqref{eq-Uprime}.

\subsection{Kinetic Flux ($F_{\rm kin}$)}

The vertical component of the kinetic flux in three dimensions is defined by

\begin{equation}
  F_{\rm kin} = \frac{1}{2} \langle \rho u'' \left( u''^2 + v''^2 + w''^2 \right) \rangle.\label{eqFkin}
\end{equation}

We define this flux with the primed velocities to remove short-lived trends and 
to make the fluxes meaningful even when computed over short time intervals. When the 
model is statistically stationary and the averaging interval is long enough, these 
fluxes coincide with those defined in \citet{NordlundStein2001}.
Furthermore, the viscous flux is defined as the horizontal average of the viscous stress tensor
as in equation (16) from \citet{NordlundStein2001} or equation (7b) in \citet{Canuto1997b}.

\end{document}